\documentclass{emulateapj}
\usepackage{natbib}
\usepackage{epsf}
\newcommand{\tx}[1]{\textrm{#1}}

\newcommand{\kms}{km~$\tx{s}^{-1}$}

\newenvironment{inlinefigure}{
\def\@captype{figure}
\noindent\begin{minipage}{0.999\linewidth}\begin{center}}
{\end{center}\end{minipage}\smallskip}

\shorttitle{Intracluster Supernovae MENeaCS}
\shortauthors{Sand et al.}

\begin{document}
 \title{Intracluster supernovae in the Multi-Epoch Nearby Cluster Survey}

\author{David J. Sand,$\!$\altaffilmark{1,2,3} Melissa
L. Graham,$\!$\altaffilmark{2,3,4} Chris Bildfell,$\!$\altaffilmark{4}
Ryan J. Foley,$\!$\altaffilmark{5,6} Chris Pritchet,$\!$\altaffilmark{4}
Dennis Zaritsky,$\!$\altaffilmark{7} Henk Hoekstra,$\!$\altaffilmark{8}
Dennis W. Just,$\!$\altaffilmark{7} St\'{e}phane Herbert-Fort,$\!$\altaffilmark{7} Suresh Sivanandam$\!$\altaffilmark{7}} \email{dsand@lcogt.net}

\begin{abstract}

The Multi-Epoch Nearby Cluster Survey (MENeaCS) has discovered
 twenty-three cluster Type Ia supernovae (SNe) in the 58 X-ray selected
 galaxy clusters (0.05$\lesssim z \lesssim$0.15) surveyed.  Four of our
 SN~Ia events have no host galaxy on close inspection, and are likely
 intracluster SNe.  Although one of the candidates, Abell399\_3\_14\_0,
 appears to be associated in projection with the outskirts of a nearby
 red sequence galaxy, its velocity offset of $\sim1000$ km s$^{-1}$
 indicates that it is unbound and therefore an intracluster SN.  Another
 of our candidates, Abell85\_6\_08\_0, has a spectrum consistent with a
 SN 1991bg-like object, suggesting that at least some portion of
 intracluster stars belong to an old stellar population.  Deep image
 stacks at the location of the candidate intracluster SNe put upper
 limits on the luminosities of faint hosts, with $M_{r}\gtrsim-13.0$ mag
 and $M_{g}\gtrsim-12.5$ mag in all cases. For such limits, the fraction
 of the cluster luminosity in faint dwarfs below our detection limit is
 $\lesssim0.1\%$, assuming a standard cluster luminosity function.  All
 four events occurred within $\sim$600 kpc of the cluster center
 (projected), as defined by the position of the brightest cluster
 galaxy, and are more centrally concentrated than the cluster SN~Ia
 population as a whole.  After accounting for several observational
 biases that make intracluster SNe easier to discover and
 spectroscopically confirm, we calculate an intracluster stellar mass
 fraction of $0.16^{+0.13}_{-0.09}$ (68\% confidence limit) for all
 objects within $R_{200}$.  If we assume that the intracluster stellar
 population is exclusively old, and the cluster galaxies themselves have
 a mix of stellar ages, we derive an upper limit on the intracluster
 stellar mass fraction of $<0.47$ (84\% one-sided confidence limit).
 When combined with the intragroup SNe results of McGee \& Balogh, we
 confirm the declining intracluster stellar mass fraction as a function
 of halo mass reported by Gonzalez and collaborators.

\end{abstract}

\altaffiltext{1}{Harvard Center for Astrophysics and Las Cumbres Observatory Global Telescope Network Fellow}
\altaffiltext{2}{Las Cumbres Observatory Global Telescope Network, 6740
Cortona Drive, Suite 102, Santa Barbara, CA 93117, USA}
\altaffiltext{3}{Department of Physics, Broida Hall, University of
California, Santa Barbara, CA 93106, USA}
\altaffiltext{4}{Department of Physics and Astronomy, University of
Victoria, PO Box 3055, STN CSC, Victoria BC V8W 3P6, Canada}
\altaffiltext{5}{Harvard-Smithsonian Center for Astrophysics, 60 Garden
Street, Cambridge MA 02138}
\altaffiltext{6}{Clay Fellow}
\altaffiltext{7}{Steward Observatory, University of Arizona, Tucson, AZ 85721}
\altaffiltext{8}{Leiden Observatory, Leiden University, Niels Bohrweg 2, NL-2333 CA Leiden,
The Netherlands}
\altaffiltext{9}{Observations reported here were obtained at the MMT Observatory, a joint facility of the University of Arizona and the Smithsonian Institution.}

\altaffiltext{10}{Based on observations obtained with MegaPrime/MegaCam,
a joint project of CFHT and CEA/DAPNIA, at the Canada-France-Hawaii
Telescope (CFHT) which is operated by the National Research Council
(NRC) of Canada, the Institut National des Science de l\'Univers of the
Centre National de la Recherche Scientifique (CNRS) of France, and the
University of Hawaii.}

\altaffiltext{11}{Based on observations obtained at the Gemini
Observatory (under program IDs GN-2009A-Q-10 and GN-2008B-Q-3), which is
operated by the Association of Universities for Research in Astronomy,
Inc., under a cooperative agreement with the NSF on behalf of the Gemini
partnership: the National Science Foundation (United States), the
Science and Technology Facilities Council (United Kingdom), the National
Research Council (Canada), CONICYT (Chile), the Australian Research
Council (Australia), Ministerio da Ciencia e Tecnologia (Brazil) and
Ministerio de Ciencia, Tecnologia e Innovacion Productiva (Argentina).}

\section{Introduction}

Galaxy clusters contain a population of stars gravitationally unbound to
individual galaxies, yet still bound to the clusters' overall
gravitational potential, created by the stripping of stars from galaxies
during interactions and mergers.  Much work has been done to
characterize the distribution, stellar mass and stellar population of
the intracluster light (ICL) through a variety of observational tracers
\citep[e.g.][among
others]{Galyam03,Arnaboldi04,Zibetti05,Gonzalez05,Krick07,Williams07,Rudick10}
because it is an important remnant of the hierarchical accretion history
of clusters and the environmental processes which can transform cluster
galaxies.  Numerical work has sought to reproduce the observed results,
to understand the processes that form the ICL, and how those processes
in turn affect cluster galaxies and the cluster as a whole
\citep[e.g.][among
others]{Murante04,Willman04,Sommer-Larsen05,Purcell08,Puchwein10,Dolag10}.

One independent and complementary technique for probing the ICL are
hostless, IC SNe.  For instance, \citet{Sivanandam09} have shown that,
under reasonable assumptions, IC SNe can generate $\sim$30\% of the iron
observed in the IC medium out to $r_{500}$, although this is subject to
the still-large uncertainties in the SN Ia rate in galaxy clusters
(especially in the relative rates of hostless and hosted SNe) and the
evolution of the ICL with respect to the cluster as a whole. More
relevant for the current work, the relative numbers of hostless and
hosted SNe Ia have been used as a measure of the IC stellar mass
fraction \citep[e.g.][]{Galyam03,McGee10}.  Additionally, hostless SNe can
serve as a tracer of the extent and distribution of the ICL -- even the
most diffuse and low surface brightness elements of the ICL can be
revealed (albeit with relative rarity) if a SN explodes in its midst.
As IC SNe observations become more common and better studied, the
relative mix of core collapse SNe and SNe Ia can constrain the age of
the underlying stellar population.  Indeed, even the relative mix of the
known sub-classes of SNe Ia -- where SN~1991T-like events (and others
with slow-declining light curves) are associated with young stellar
populations \citep[e.g.][]{Hamuy95}, while SN~1991bg-like events (and
those with fast-declining light curves) are associated with old ones
\citep[e.g.][]{Howell01}~ -- can serve as a tracer of the underlying
stellar population.

Intracluster SN observations are still rare.  A first possible
IC SN Ia was discovered by \citet{Smith81}, in an apparent
stream connecting NGC~4406 and NGC~4374 in the Virgo cluster.
\citet{Galyam03} found two IC SNe out of a total sample
of seven in the WOOTS survey, a cluster SN search centered at $z\sim
0.1$ \citep{Sharon07,Galyam08}.  More recently, \citet{McGee10} mined
the Sloan Digital Sky Survey (SDSS) SN Survey for SNe Ia
associated with low redshift galaxy groups.  Of their 59 group SNe Ia,
19 have no detectable host galaxy, and after making corrections for the
group luminosity function and assuming a two-component SNe Ia rate
\citep[e.g.][]{Mannucci05,Scannapieco05} they found that
$47^{+16}_{-15}$\% of the stellar mass in their groups was in the form
of diffuse light.  Additionally, \citet{Dilday10} discovered up to three
hostless, IC SNe Ia in their study of the SNe Ia rate in
galaxy clusters with the SDSS-II SN Survey.  Several other
possible IC SNe have been discovered over the years
\citep[e.g.][]{GermanyIC,ReissIC,SandSN08,GalyamIC2,Sharon10,Barbary10}.  As
outlined in the previous paragraph, hostless SNe are a unique
probe of the diffuse universe.  Much more can be done, and there is a
need for systematic surveys for hostless SNe over the entire
range of halo masses to study the spatial extent, stellar content and
mass of diffuse light at all scales.

We have recently completed the Multi-Epoch Nearby Cluster Survey
(MENeaCS), one of whose many goals is to measure the SN~Ia rate in the
most massive, X-ray selected $z\sim0.1$ galaxy clusters.  In this paper,
we report the discovery of four likely IC SN~Ia which we have
spectroscopically confirmed from among a sample of twenty-three in the
MENeaCS survey. We use these to constrain the IC stellar population and
stellar mass fraction in the MENeaCS cluster sample, along with the
radial distribution of IC stars.  The definition of ICL is necessarily
ambiguous, given that galaxies have no well-defined 'edge'
\citep[e.g.][]{Abadi06}.  We thus adopt the point of view of
\citet{Rudick10} and others: the ICL includes all low luminosity
components of galaxy clusters, whether it is large scale diffuse
luminosity, a tidal stream, or even the extreme outskirts of a cluster
member.  Where necessary, we adopt a flat cosmology with $H_{0}$=70 \kms
Mpc$^{-1}$, $\Omega_{m}$=0.3, and $\Omega_{\Lambda}$=0.7.

\section{Survey Description}\label{sec:surveydesc}

While a full description of the MENeaCS survey, along with our cluster
SNe rates, will be presented in a separate paper (Sand et al. in prep),
we briefly describe the relevant aspects of the survey here.  The
principal time-domain goal of MENeaCS is to measure the cluster SN~Ia
rate at $z\sim0.1$, including both hosted and hostless, IC events.  We
take advantage of CFHT's queue scheduling to observe $\sim$30 clusters
per queue run (with 2 $\times$ 120 s $g'$ and $r'$ images), and a total
of 58 unique clusters throughout the year.  These clusters were X-ray
selected, and have $0.05 \lesssim z \lesssim 0.15$.  The monthly CFHT
imaging is supplemented by follow-up imaging with the 2.3~m Bok
telescope on Kitt Peak with its 90Prime imager \citep{90prime}.  Viable
cluster SNe~Ia are identified based on brightness and color selection,
and spectroscopically confirmed via monthly spectroscopy, primarily with
the MMT. Target of Opportunity (ToO) observations with Gemini were
acquired specifically to follow up our hostless SN candidates, which
were pursued without any prior cut on brightness or color.  The survey
began in February 2008 and spectroscopic observations ended in December
2009.  The last of the imaging data was taken in 2010A in order to get
the image depth necessary for the other scientific aspects of the
MENeaCS program.

During the course of the survey, we maintained deep-stacked images of
all of our survey fields -- consisting of all previous images at that
position -- for the purposes of faint host identification.  For every SN
candidate we attempt to identify a host galaxy using the method of the
SN Legacy Survey \citep{Sullivan06}, which uses a host-SN elliptical
separation normalized by the candidate host galaxy size to assign hosts
-- the dimensionless $R$ parameter described in Eqn.~1 of Sullivan et
al. (2006).  Each host is characterized by an elliptical shape, as
output by SExtractor, with semimajor and semiminor axes ($r_A$ and
$r_B$) along with a position angle, $\theta$.  We note here that Eqn.~1
of Sullivan et al. (2006) had a typographical error. The corrected
version of the formula -- where ($x_{\mathrm{SN}}$,$y_{\mathrm{SN}}$) is
the SN pixel position and ($x_{\mathrm{gal}}$,$y_{\mathrm{gal}}$) is the
host center -- should read

\begin{equation}
R^2={C_{xx}}x_r^2 + {C_{yy}}y_r^2 + {C_{xy}}x_ry_r,
\end{equation}

\noindent where $x_r=x_{\mathrm{SN}}-x_{\mathrm{gal}}$,
$y_r=y_{\mathrm{SN}}-y_{\mathrm{gal}}$,
$C_{xx}=\cos^2(\theta)/r_A^2+\sin^2(\theta)/r_B^2$,
$C_{yy}=\sin^2(\theta)/r_A^2+\cos^2(\theta)/r_B^2$, and
$C_{xy}=2\cos(\theta)\sin(\theta)(1/r_A^2-1/r_B^2)$. 

Unfortunately, the original typographic error propagated into our survey
code, although as we shall see from Figure~1, it did not ultimately
effect our ability to distinguish between hosted and hostless SNe.
Typically, the apparent detectable edge of a galaxy is at $R\sim3$,
although this is clearly a depth dependent statement.  As in the SNLS,
we initially classify a SN candidate as hostless if it has no potential
hosts with $R<5$.  These candidate IC SNe are flagged and a visual
search for a faint host is performed with our deep-stacked image, in
case it was missed by our automated analysis.  Very occasionally, the
visual search revealed that a SN candidate with $R>5$ did seem to be
associated with the outer halo of a galaxy (in retrospect, this was due
to the typographical error mentioned above), and the SN was followed up
during our classical spectropscopy observations.  The individual
discovery $g'$ and $r'$ band images are blinked and examined by a human
to verify that the IC SN candidate is not a slowly moving object.  If
the candidate is still viable, we triggered our Gemini ToO program, or
otherwise placed the hostless SN candidate into our 'high priority'
spectroscopic follow up list.  Given our deep imaging (see
\S~\ref{deepsec}), our adopted definition of a hostless SN should be
comparable to or stricter than recent IC SN work, such as
\citet{McGee10}, who used twice the radius at which 90\% of the light
was enclosed (down to SDSS depths) as their working definition of a
hostless SN.  We also point out that, given the stacking of our images
prior to SN discovery (with our shallowest stacks reaching
$M_{g}\sim-12.5$ mag; $M_{r}\sim-13.0$ mag-- see \S~\ref{sec:hostlim}),
our host galaxy limits are stricter than that of \citet{McGee10}, which
were at best $M_{r}\sim-15.0$ mag, and similar to \citet{Galyam03}.

In general, our classically scheduled spectroscopic follow up was
focused on likely cluster SN~Ia.  If a SN candidate was newly
discovered, its $g-r$ color and magnitude had to be consistent with a SN
Ia within $\sim$30 days of explosion at the cluster redshift, which
ranged between $-0.5 \lesssim g - r \lesssim 0.8$ and $ 17.5 \lesssim g
\lesssim 22.5 $ (we have calibrated all survey imaging to the SDSS
filter system; \S~\ref{sec:phot}) and had to have a clustercentric
distance of $R<1.2R_{200}$ (we have derived $R_{200}$ for each of the
MENeaCS clusters using the X-ray scaling relations of
\citet{Reiprich02}, see \S~\ref{sec:ICpop}).  Although every effort was
made to spectroscopically confirm viable SNe Ia at higher clustercentric
radius, this was not always possible.  We also attempt to get
spectroscopic confirmation of SNe even if they appear over 1 month old,
based on our cadence, weather on previous spectroscopic runs, and the
light curve of the SN candidate.  Note that all viable cluster SN~Ia
candidates, as defined above, were observed modulo telescope scheduling
and weather.  Other SN candidates (i.e. those that were not consistent
with expected cluster SN Ia's colors or luminosity, but still with
$g\lesssim22.5$) were followed up with lower priority, but this yielded
7 cluster core collapse SNe.  These allow us to set a limit on the
number of hostless to hosted core collapse SNe in the cluster
environment for the first time (\S~\ref{sec:ICpop}), a notion
particularly relevant now that some star formation has been observed in
the IC environment \citep{Sun07,Sun10,Sivanandam10,Smith10}, and has
even been suggested by simulations \citep{Puchwein10}.

MENeaCS has spectroscopically confirmed twenty-three cluster SN Ia (all within
3000 km s$^{-1}$ of the cluster redshift), four of which have no host
upon close inspection. We plot the SN-to-host distance for our cluster
SN Ia sample in Figure~\ref{fig:host_SN_dist}, in terms of their
effective radius, $R_{e}$, after host assignment via the elliptical
radius parameter as described above.  The effective radius of each host
galaxy is estimated via the {\sc flux\_radius} parameter in SExtractor
\citep{sexbib} (this parameter was set so that it reported the radius at
which 50\% of the galaxy light was enclosed, the traditional definition
of effective radius).  The distances, in effective radii, of our four
hostless cluster SNe Ia with respect to the nearest source in our
SExtractor catalogs are shown as dashed lines (see \S~\ref{deepsec} for
our limits on faint hosts associated with our IC SN candidates).  Note
that the SExtractor {\sc flux\_radius} parameter does not account for
the point spread function (PSF) width, and so our $R_{e}$ values will be
overestimates, with the corresponding host -- SN distance becoming
larger.  This plot reinforces the distinct nature of our IC SN
candidates.

\section{Observations \& Analysis}

In this section, we will present the observations of our IC
SNe along with our evidence for them not being related to a visible
host.  In all, we have identified three cluster SN Ia which are
excellent candidates for having progenitors which are IC
stars, along with an intriguing SN which may be associated with debris
or other stripped material from a nearby cluster galaxy. We summarize
the observations of each SN in Figures~2--5.  The results of our
spectroscopic analysis are summarized in Table~\ref{table:spectable},
and our magnitude limits for faint hosts are in
Table~\ref{table:deeptable}.

\subsection{Discovery and follow up photometry}\label{sec:phot}

We have implemented a nearly real time analysis system for discovering
and flux calibrating SN candidates in our CFHT imaging.  This
pipeline, as discussed in \S~\ref{sec:surveydesc}, also attempts to
assign a host to every discovered transient.  If no host is found, and
this is confirmed via visual inspection of the raw and processed data,
the SN is moved to the top of our spectroscopic priority list, and our
Gemini ToO was triggered, if available.  Since our imaging consisted 
of not only our CFHT/Megacam data, but also 2.3 m Bok and 
occasional Gemini ToO acquisition images, we chose to calibrate our images to
the Sloan Digital Sky Survey $g,r$ photometric system, which in
general required both a zeropoint and color term.

\subsection{Spectroscopic confirmation}

We used our Gemini Observatory ToO program to confirm three of our
candidate IC events using the Gemini-North Multi Object Spectrograph
\citep{Hook04}.  Long-slit spectra were taken with the R400 grating
centered at 680 nm, with the GG455 order blocking filter, covering
$\sim$465-890 nm.  The 1\farcs0 slit was used and was oriented so that a
relatively bright star was also placed in the slit, in case the SN was
not visible in acquisition images.  In the case of Abell\_399\_3\_14\_0
(note that we adopt our internal MENeaCS identification for our SNe,
although we list IAU circular identifiers when available in Table~1),
the slit was oriented so that a redshift for the nearby cluster red
sequence galaxy could be obtained simultaneously.  All SN spectra were
reduced in a standard manner utilizing {\sc IRAF}\footnote{IRAF is distributed by the National Optical Astronomy Observatory, which is operated by the Association of Universities for Research in Astronomy (AURA) under cooperative agreement with the National Science Foundation.} routines, including
the Gemini-specific {\sc IRAF} package.

We have also used Hectospec on the MMT \citep{hectospec} in a queue
mode to obtain a spectrum of Abell1650\_9\_13\_0. Hectospec is a 300
fiber (1\farcs5 diameter) spectrograph with a 1 square degree field of
view.  The 270 line grating was in place, giving wavelength coverage
between $\sim$360-800 nm.  The spectrum was pipeline processed at the
Harvard-Smithsonian Center for Astrophysics.

The spectra of our IC SNe can be seen in Figures~2--5.  We
utilize the Supernova Identification (SNID) software package
\citep{SNID} for classifying all of our spectra, which cross-correlates
the input spectrum with a library of template SN spectra to both
classify and determine the redshift of our SNe, based on the
technique of \citet{Tonry79}.  We mimic the four step procedure of
\citet{Foley09} for determining a SN's type, subtype, redshift
and age using SNID.  The reader is referred to that work for the details
of the process, and can see our results for the MENeaCS IC
SNe in Table~\ref{table:spectable}.  The best-matched SNID
template spectrum is also listed in Table~\ref{table:spectable}, and is
overplotted in Figures~2--5 for comparison with the observed spectrum.
In general, the match between template and spectrum is excellent, and we
discuss the individual fits in \S~\ref{sec:note_ind}.

\subsection{Deep stack images}\label{deepsec}

With the survey complete, we have stacked all CFHT imaging epochs
without SN contamination to search for faint hosts in the
vicinity of our IC SNe.  In general, this consisted of
imaging taken prior to SN discovery only; close examination of
images taken a year or more after SN discovery showed some residual SN
light.  Our final deep stack catalog for each field was made by median
combining these SN-free images, using the {\sc
SWarp}\footnote{version 2.15.7; http://terapix.iap.fr/soft/swarp}
software package, and then running SExtractor on the resulting image.
Since the goal was to identify the faintest possible hosts, we set the
SExtractor keywords {\sc Detect\_minarea} to 5, {\sc Detect\_thresh} to
2, and {\sc Analysis\_thresh} to 2.  No source was found within $R<5$
(see \S~\ref{sec:surveydesc}) confirming our original trigger on these
events as potential IC SNe.  We visually inspected the
IC SN positions as well, again finding no host.

To quantify the depth of these deep stack images, we implanted
artificial stars ($\sim$50000 per image, over several iterations) with a
PSF matched to that of the original image.  We note that dwarf galaxies
in the Coma cluster have a typical size of $\sim$1 arcsecond
\citep[e.g.][]{Komiyama02}.  Given that our cluster sample is at higher
redshift, and has a typical seeing of $\sim$0.8-1.0 arcseconds, our
decision to quantify our sensitivity to faint hosts via a standard
point source detection efficiency calculation is justified.  The $g$
magnitude of the artificial stars was drawn randomly from 21 to 28 mag,
with uniform probability.  The $g-r$ color was then randomly assigned
over the range $0.5 < g-r < 1.5$, again with uniform probability.  No
artificial star was put within ten $R_{e}$ of any source in our original
SExtractor catalog, as determined via the {\sc FLUX\_RADIUS} output
parameter.  Note that our statistics for sources $\gtrsim$5$R_{e}$ -- more
representative of our actual IC SN population as seen in Figure~1 --
and $>$10$R_e$ are identical (see \S~4.1).  After implantation, our
detection efficiency as a function of magnitude was calculated by
re-running SExtractor on the image.  We adopt the magnitude at which we
recover 50\% of our artificial sources as our image depth, and we report
this in Table~\ref{table:deeptable}, after translating into absolute
magnitudes given the redshift of each cluster.  We inspected several
images to confirm that the 50\% recovery threshold in our SExtractor
catalog was visually correct.  We calculate what fraction of the
clusters' luminosity is contained in dwarf galaxies below our detection
limit, and the implications for that on the hostless nature of our IC
SN candidates, in \S~\ref{sec:hostlim}.

\subsection{Notes on individual objects}\label{sec:note_ind}

Here we present additional details for each of the IC SN candidates.

\subsubsection{Abell1650\_9\_13\_0}

\begin{figure}
\begin{center}
\mbox{ 
 \epsfysize=4.0cm \epsfbox{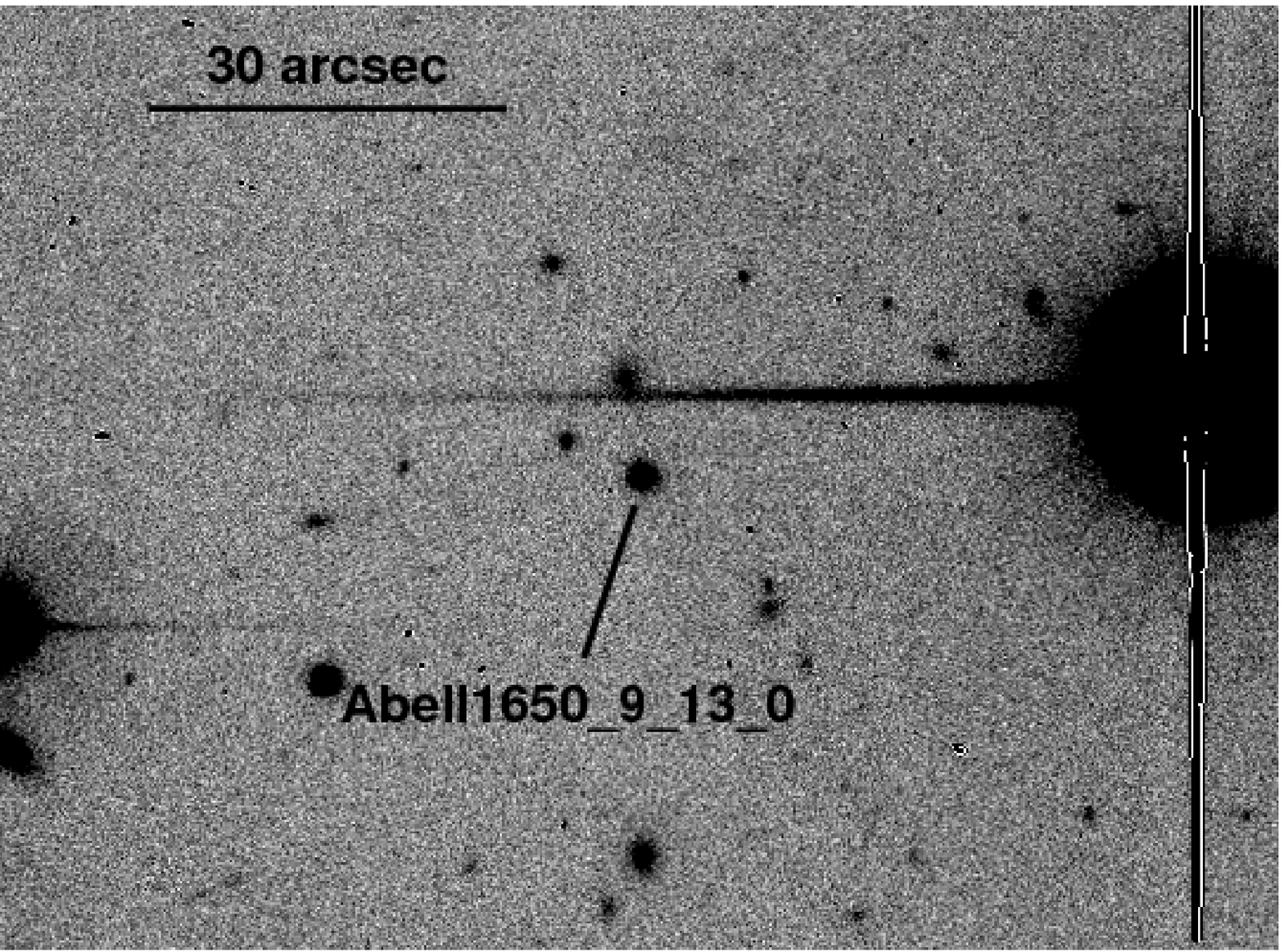} 
 \epsfysize=4.0cm \epsfbox{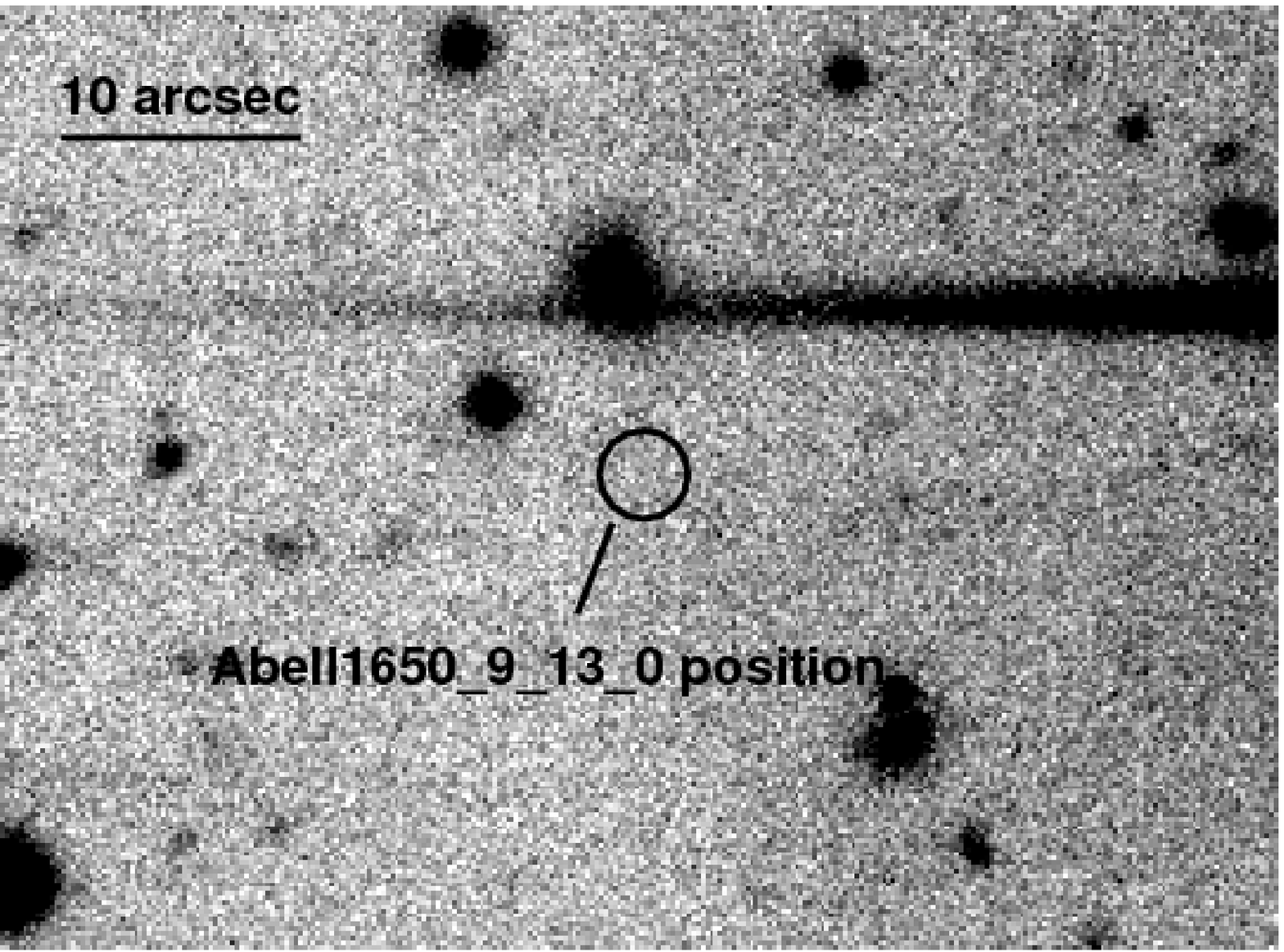} 
 \epsfysize=4.0cm \epsfbox{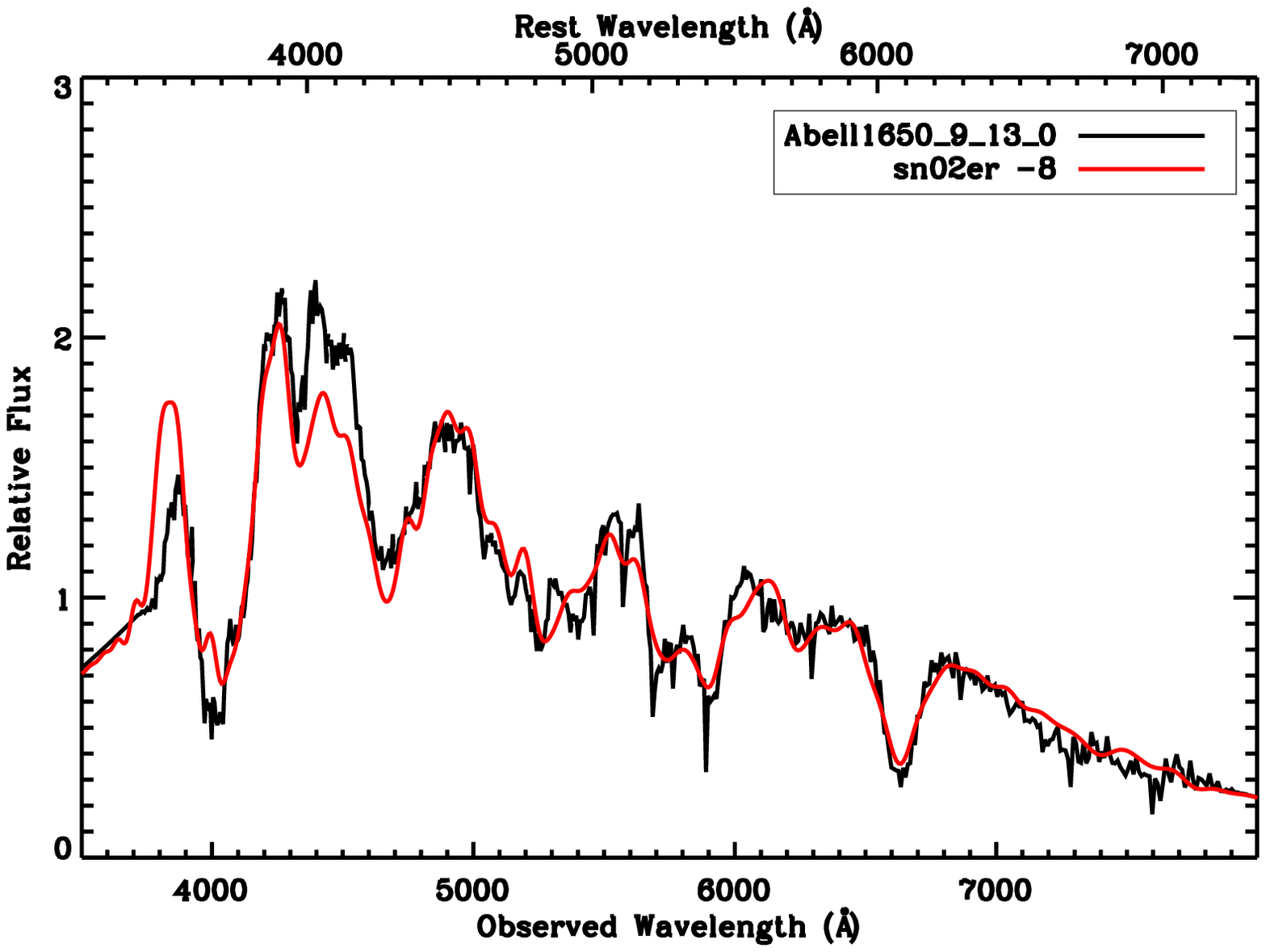}
} 
\caption{Left: Discovery $r'$ image of Abell1650\_9\_13\_0, which
is located $\sim$470 kpc East of the BCG in Abell 1650 ($z=0.0838$).
North is up and East is to the left.  Middle: A zoom in of the SN
location on our deep stack, SN-free image.  No host is apparent down
to $M_{g}=-12.47$ mag, $M_{r}=-13.04$ mag at the redshift of the cluster
(Table~\ref{table:deeptable}).  Right: MMT/Hectospec spectrum of
Abell1650\_9\_13\_0, along with the best-fitting SNID template,
SN2002er, which is overplotted in red.  The SN is a normal SN Ia
at $z=0.0836$ (see Table~\ref{table:spectable}).
\label{fig:SNA1650}}
\end{center}
\end{figure}

Abell1650\_9\_13\_0 was discovered 2009 December 14 (UT dates are used
throughout this paper) with $g=18.3$ mag and $r=18.6$ mag, making it a
prime target for spectroscopic follow up.  A spectrum was taken with
MMT/Hectospec on 2009 December 21, which indicates a normal SN Ia near
maximum light (Figure ~\ref{fig:SNA1650}).

The SN position is $\sim$470 kpc East of Abell 1650's BCG.  Deep stack
images of Abell 1650 with no SN contamination indicate a depth of
$M_{g}\sim-12.5$ mag and $M_{r}\sim-13.0$ mag at the cluster redshift
($z=0.0838$).  The SN is $\sim$10 effective radii from the nearest
object in our deep Abell 1650 catalog.

\subsubsection{Abell2495\_5\_13\_0}

\begin{figure}
\begin{center}
\mbox{ \epsfysize=4.0cm \epsfbox{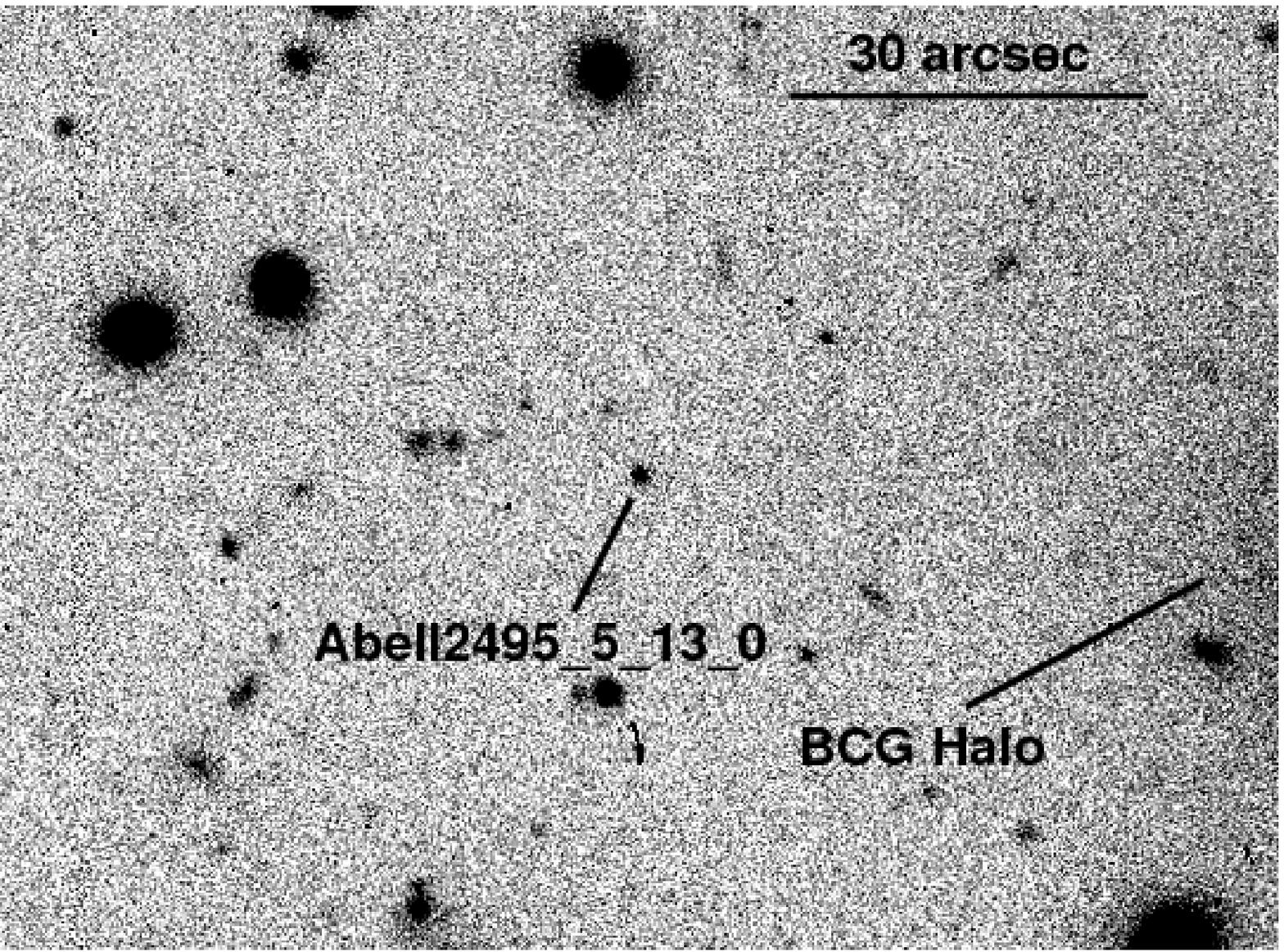} \epsfysize=4.0cm
\epsfbox{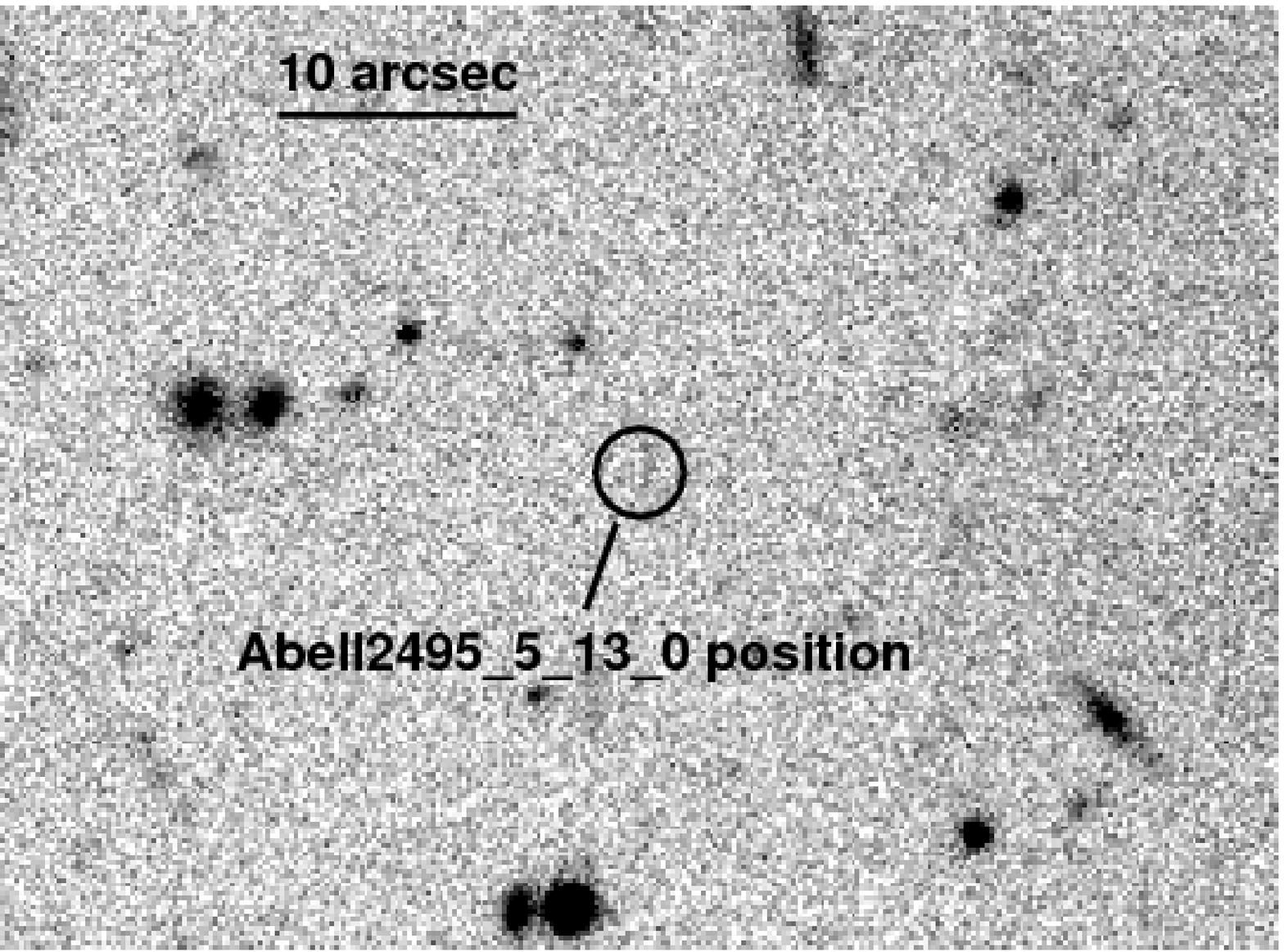} \epsfysize=4.0cm \epsfbox{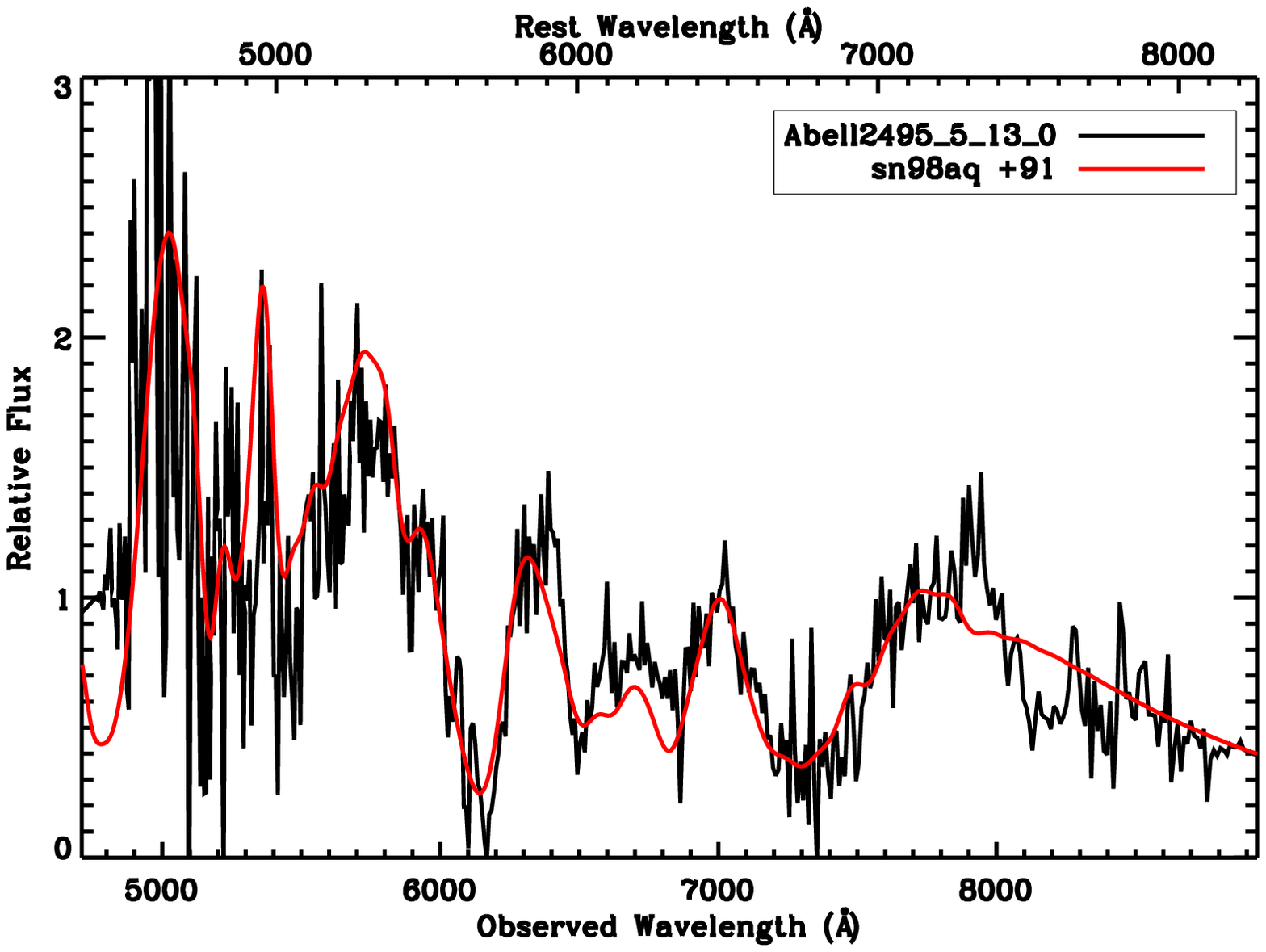}}

\caption{ Left: Discovery $r'$ image of Abell2495\_5\_13\_0, which
is located $\sim$150 kpc Northeast of the BCG in Abell~2495
($z=0.0775$).  Middle: A zoomed in view of the SN location in our
deep stack, SN-free image.  No host is apparent down to $M_{g}=-11.72,
M_{r}=-12.37$ mag.  Right: Gemini/GMOS spectrum of Abell2495\_5\_13\_0,
along with the best-fitting SNID template, SN1998aq, which is
overplotted in red.
\label{fig:SNA2495}}
\end{center}
\end{figure}

Abell2495\_5\_13\_0 was discovered on 2009 May 23 with $g=23.0$ mag and
$r=22.4$ mag, fainter than we typically followed up SNe in MENeaCS, but
due to its lack of an apparent host and the fact that Abell 2495 had not
been observed since October 2008, we triggered our Gemini ToO program.
Further imaging on June 17 showed that the SN was in decline, just prior
to our spectroscopic observations on June 18.  The SN spectrum indicates
that Abell2495\_5\_13\_0 is a normal SN~Ia roughly three months past
maximum light (Figure~\ref{fig:SNA2495}).

The SN position is $\sim$150 kpc from the BCG in Abell 2495, and is
the most centrally located in our sample (Figure~\ref{fig:SNA2495}).
The deep stack images indicate that there is no host galaxy at the
location of Abell2495\_5\_13\_0 to a magnitude of $M_{g}\sim-11.7$ mag and
$M_{r}\sim-12.4$ mag at the distance to Abell 2495 ($z=0.0775$).  The
nearest object in our Abell2495 deep catalog implies that the SN was
over $\sim$20 effective radii away.

\subsubsection{Abell85\_6\_08\_0}\label{A85SN_descrip}

Abell85\_6\_08\_0 was discovered on June 18, 2009 with $g=20.6$ and
$r=19.8$.  Subsequent photometry was obtained with both the CFHT and
the Steward 2.3 m, indicating a SN well past maximum light.

\begin{figure}
\begin{center}
\mbox{ \epsfysize=4.0cm \epsfbox{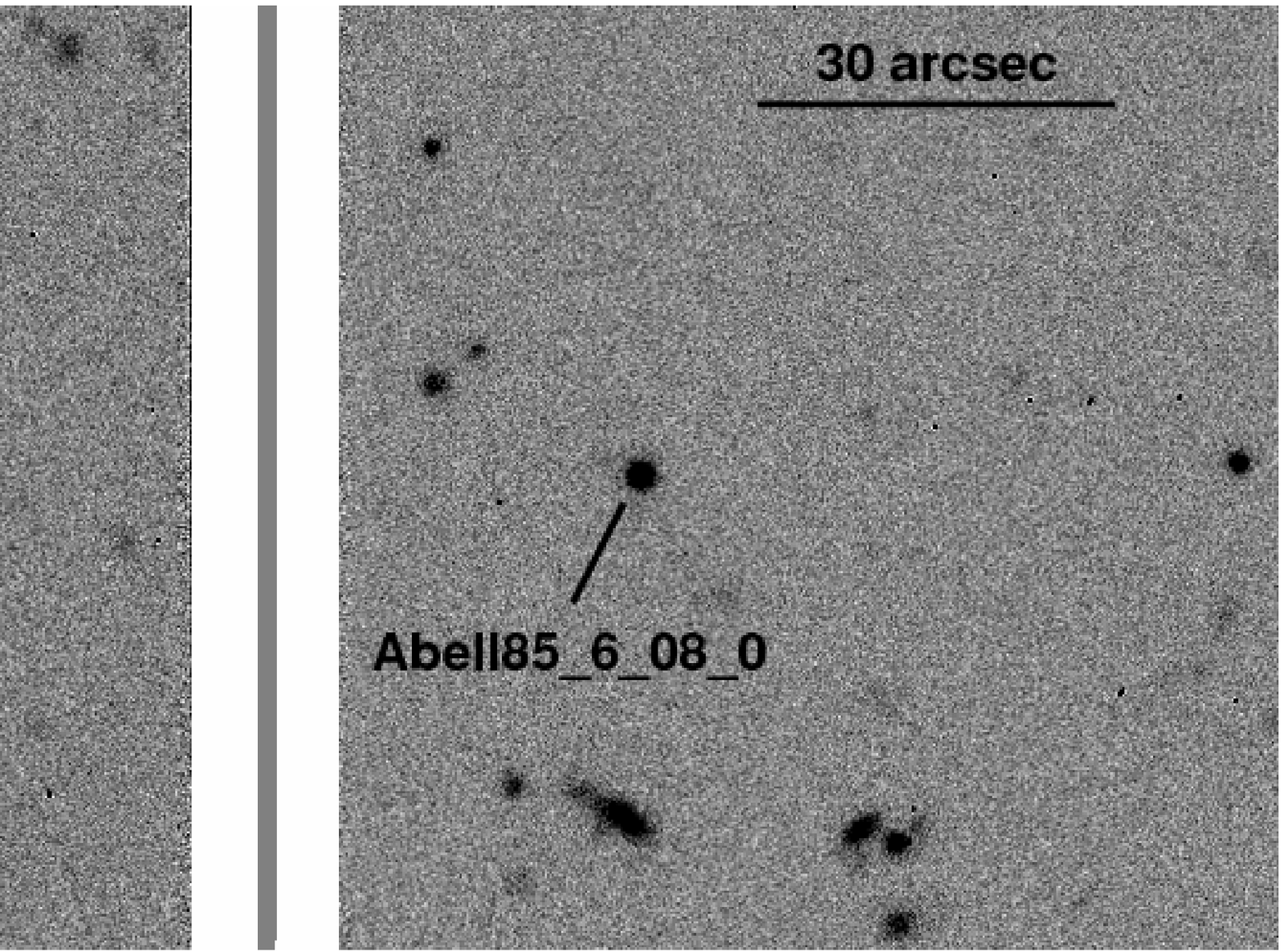} \epsfysize=4.0cm
\epsfbox{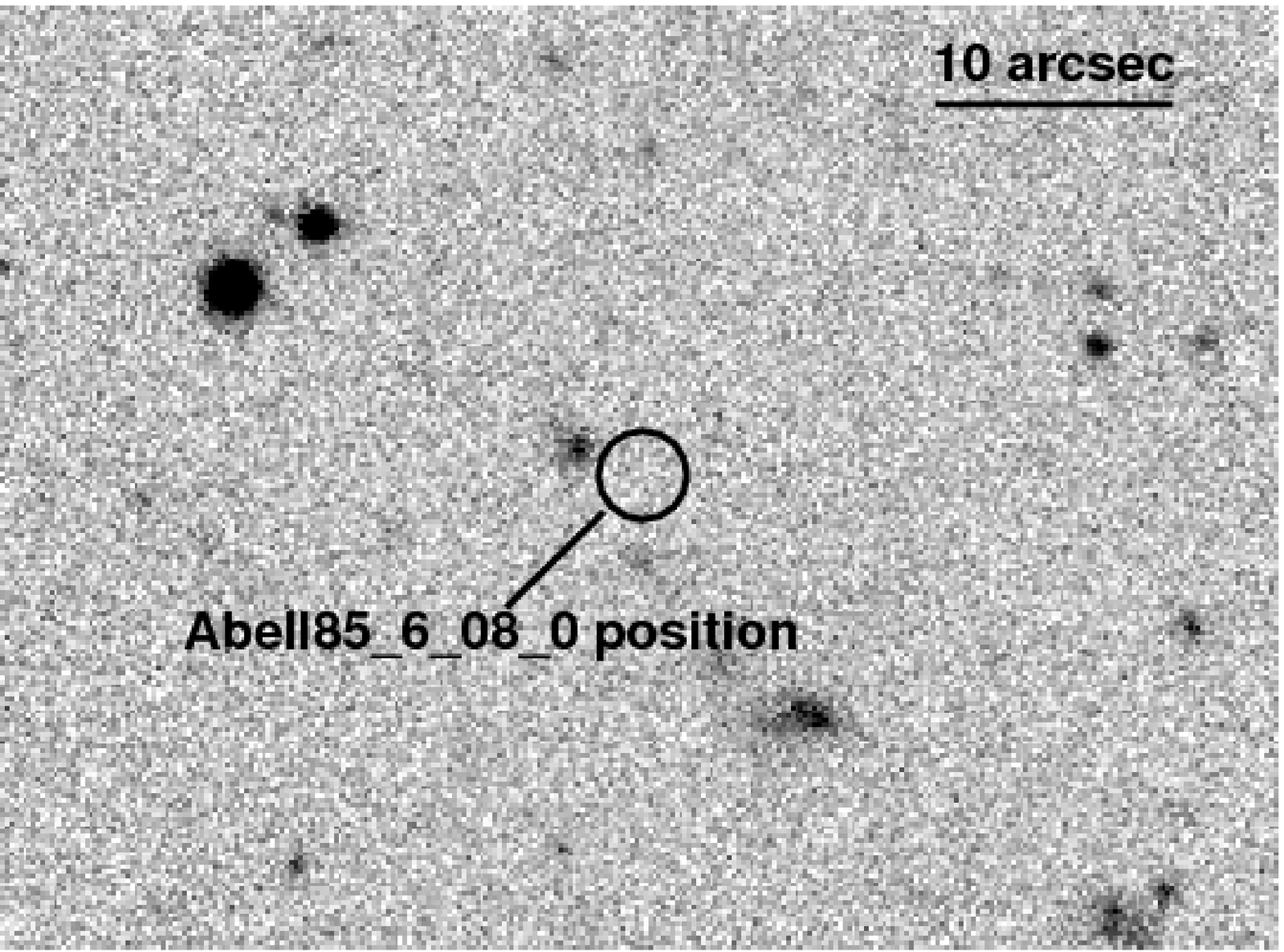} \epsfysize=4.0cm \epsfbox{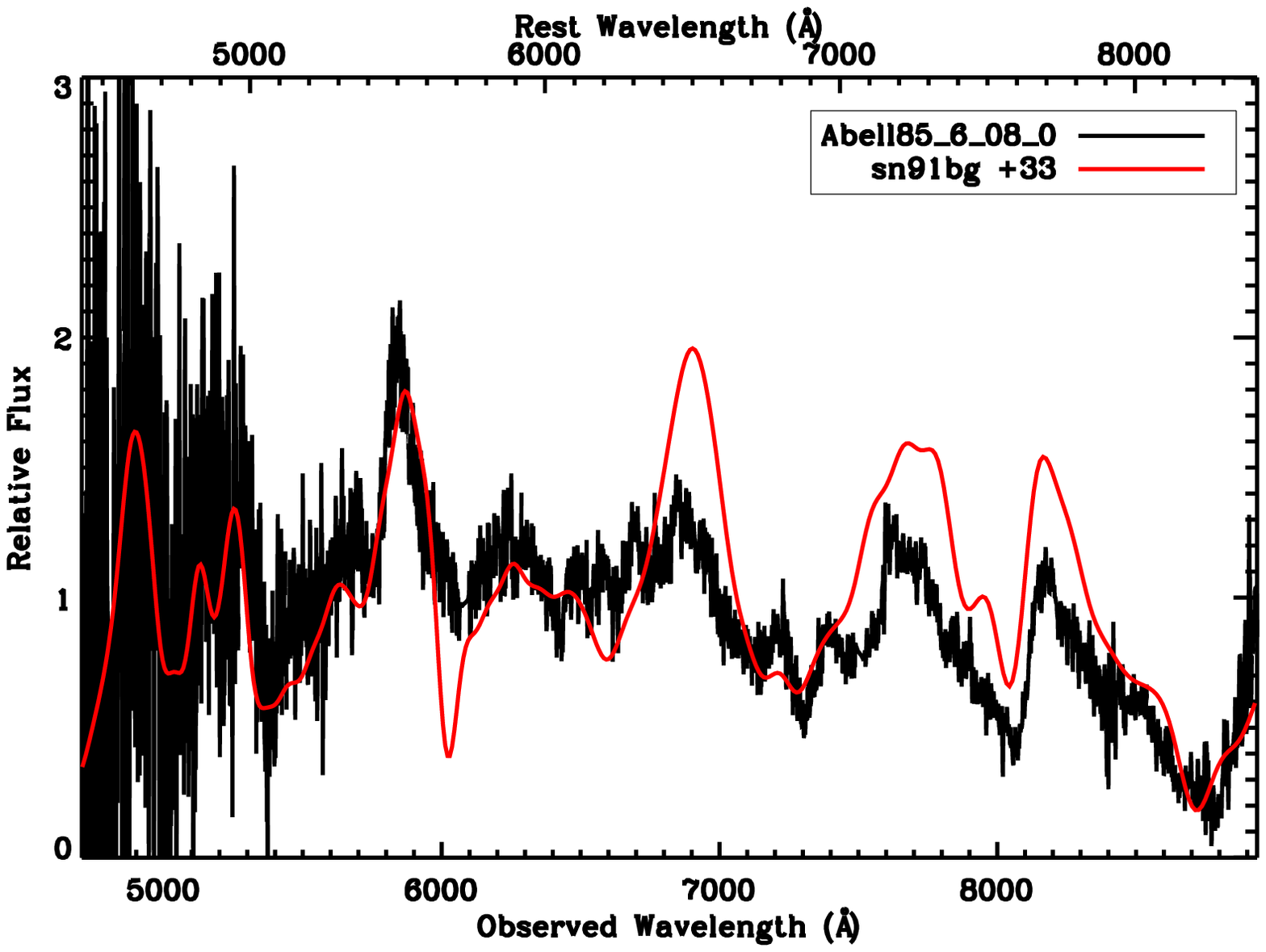}} 

\caption{Left: Discovery $r'$ image of Abell85\_6\_08\_0, which is
located $\sim$600 kpc Southeast of the BCG in Abell~85 ($z=0.0578$).
Middle: A close up view of the SN location in our deep stack, SN-free
image.  No host is apparent down to $M_{g}=-11.15, M_{r}=-11.68$ mag.
Right: Gemini/GMOS spectrum of Abell85\_6\_08\_0, along with the
best-fitting SNID template, SN91bg, which is overplotted in red.  Even
though the best-matched SNID spectrum is not as well fit as our other
spectra, we argue in \S~\ref{A85SN_descrip} that Abell85\_6\_08\_0 is
likely of the underluminous, SN~1991bg subtype.  \label{fig:SNA85}}
\end{center}
\end{figure}

The Gemini/GMOS SN spectrum is shown in Figure~\ref{fig:SNA85}.  The
best matched SNID template for Abell85\_6\_08\_0 is SN~1991bg, roughly
five weeks past maximum, and the top six SNID matches to the spectrum
are also of the SN~1991bg subtype.  We are confident that
Abell85\_6\_08\_0 is an underluminous SN Ia based on OI and Ca II in
absorption -- which is typical for underluminous SN Ia roughly 4-5 weeks
past maximum \citep[e.g.][]{Tauenberger08}.  We do not see the narrow Na
I D line near 5700 \AA, typical of underluminous SN Ia.  The SN type is
supported by the declining light curve at the time the spectrum was
taken, although no data is available near maximum light.  If an
underluminous SN Ia, Abell85\_6\_08\_0 is likely associated with an old
stellar population \citep[e.g.][]{Howell01}, which we briefly discuss in
\S~\ref{sec:ICpop}.

Abell85\_6\_08\_0 is $\sim$600 kpc southeast of Abell 85's BCG.  No
host is detected down to $M_{g}\sim-11.2$ mag and $M_{r}\sim-11.7$ mag in our
deep stack catalogs at the cluster redshift ($z=0.0578$).  The SN is
$\sim$5 effective radii from the nearest object in our deep Abell 85
catalog.

\subsubsection{Abell399\_3\_14\_0}\label{sec:A399}

\begin{figure}
\begin{center}
\mbox{ \epsfysize=4.0cm \epsfbox{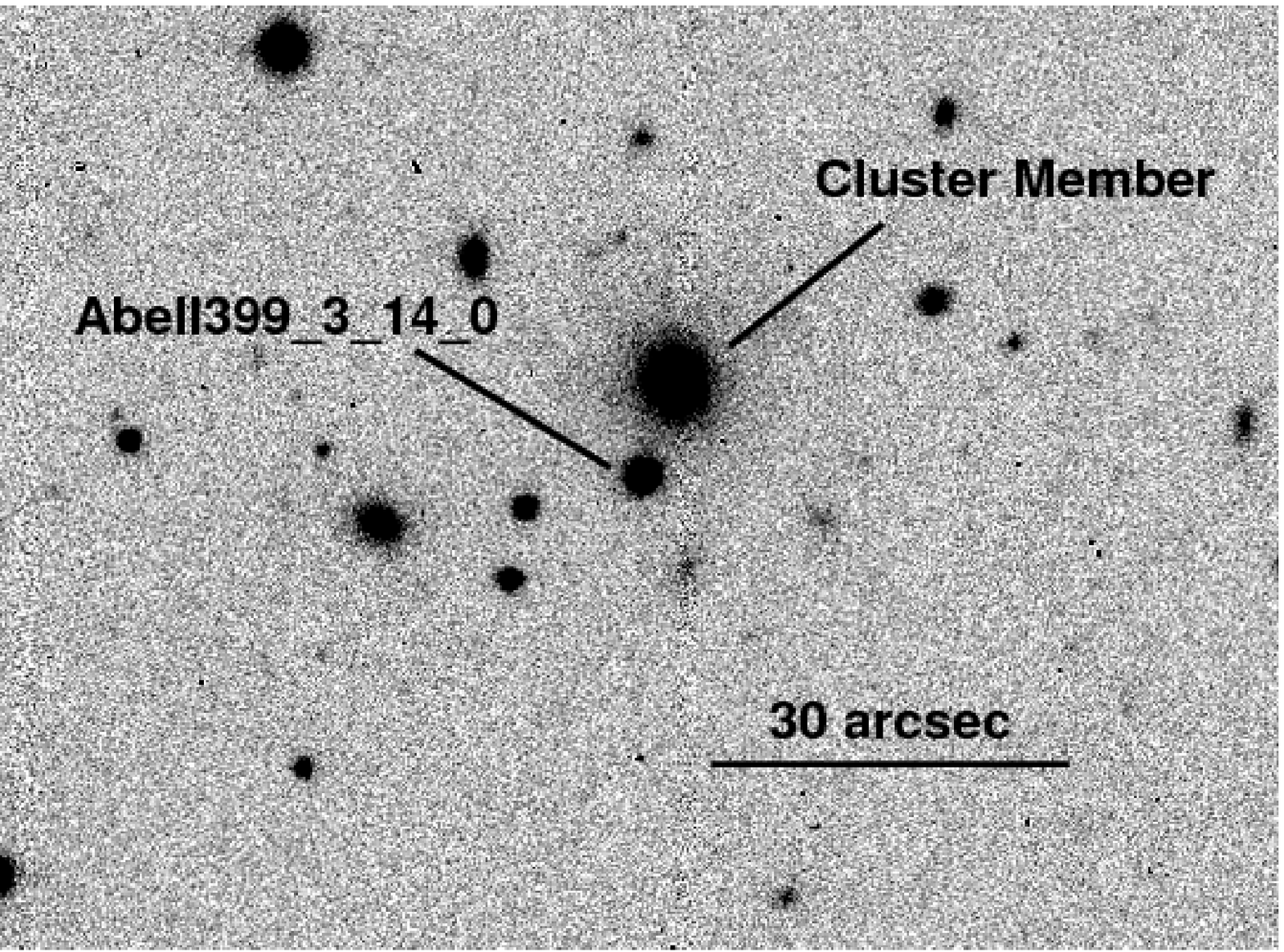} \epsfysize=4.0cm
\epsfbox{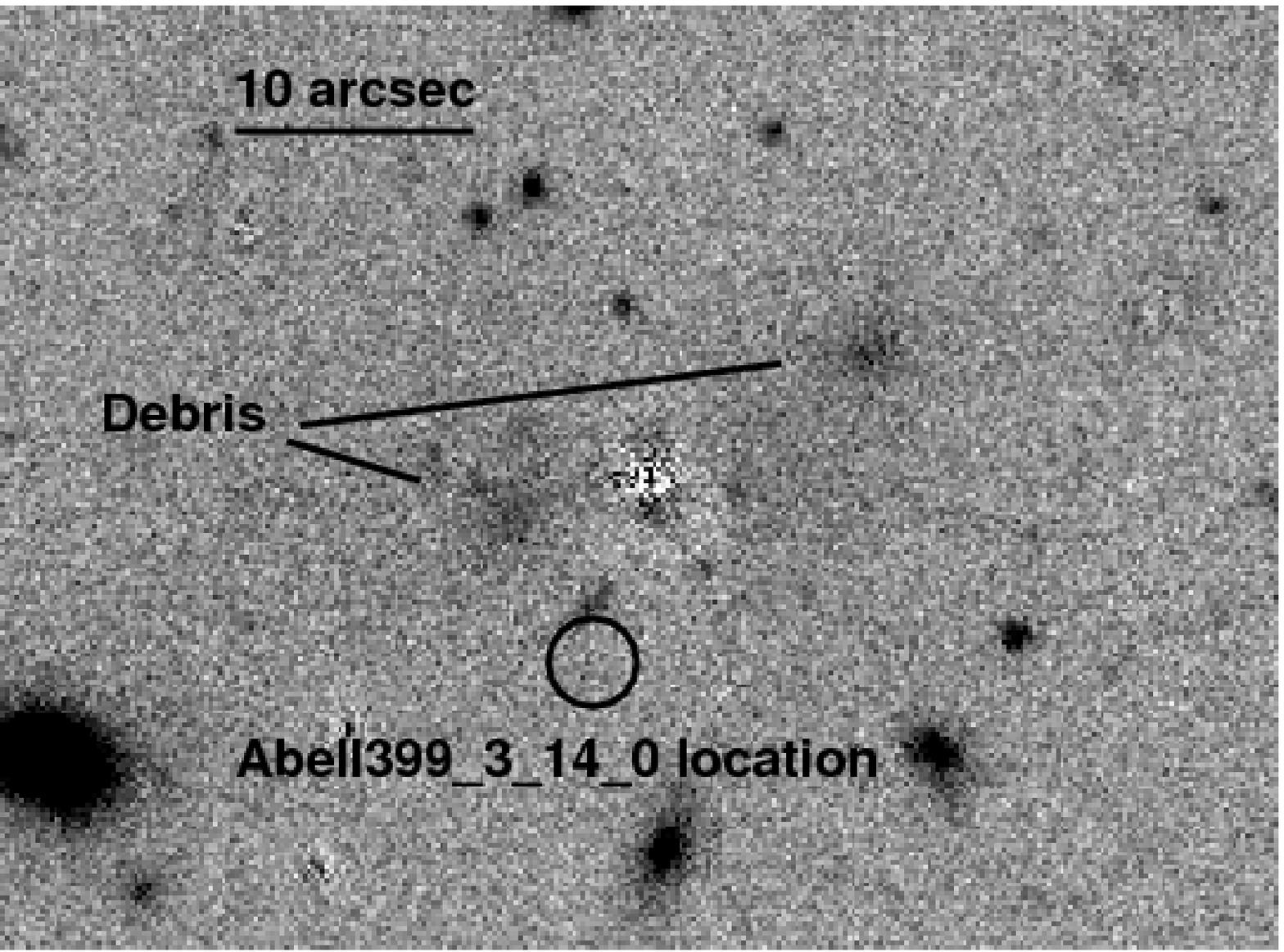} \epsfysize=4.0cm \epsfbox{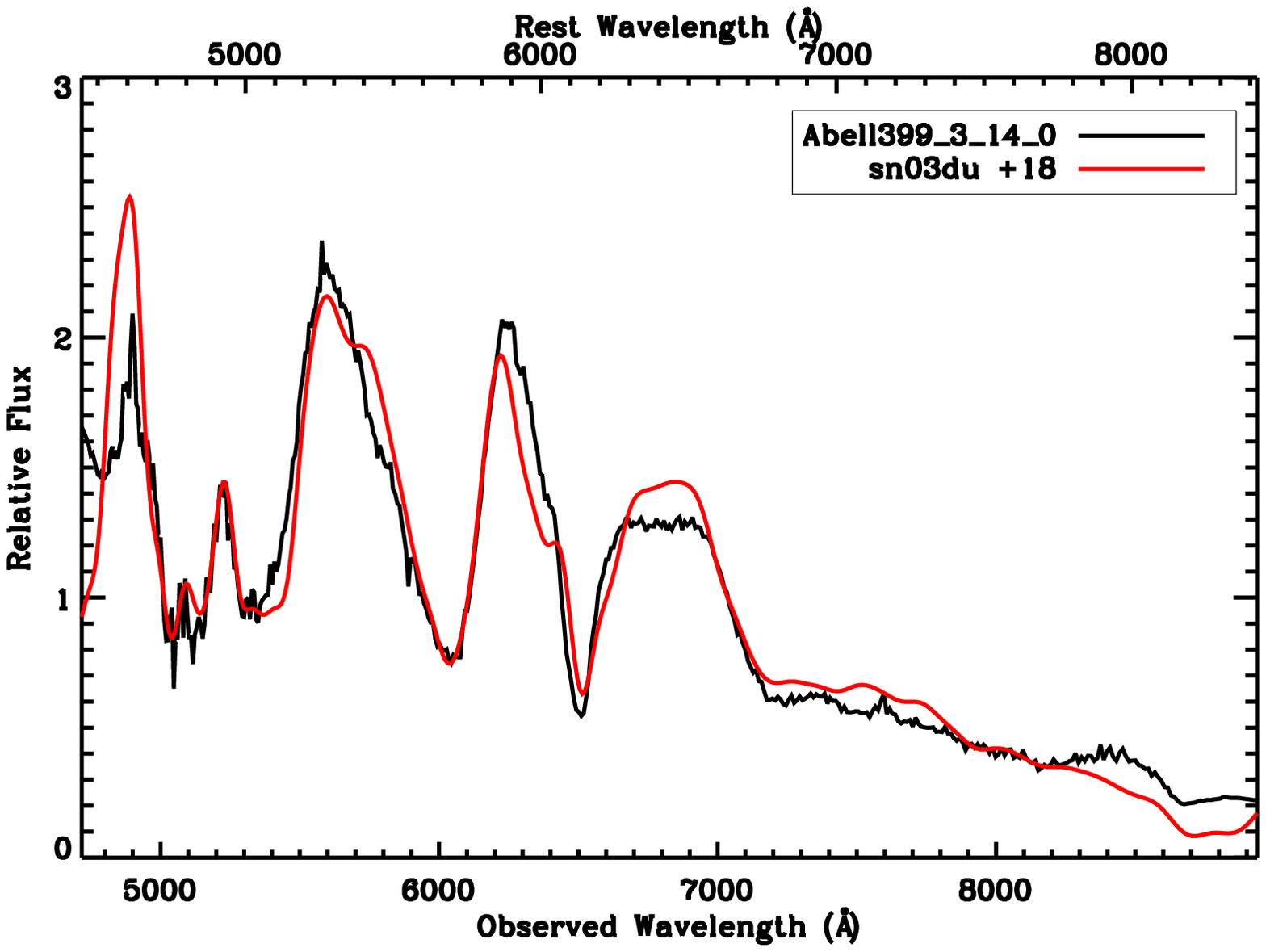}}
\caption{ Left: Discovery $r'$ image of Abell399\_3\_14\_0, which is
located $\sim$620 kpc Southwest of the BCG in Abell~399 ($z=0.0718$).
Note the relatively nearby cluster galaxy, dubbed Galaxy 1.  Middle: A
close up view of the SN location in our deep stack, SN-free image, where
the best GALFIT model of Galaxy 1 has been subtracted (along with
several other nearby galaxies).  Notice the debris uncovered after the
GALFIT model subtraction, reminiscent of that seen in local cluster
early types \citep{Janowiecki10}.  We discuss the IC status of this
object in \S~\ref{sec:A399}.  Right: Gemini/GMOS spectrum of
Abell399\_3\_14\_0, along with the best-fitting SNID template, SN2003du, a
normal type-Ia, which is overplotted in red.  \label{fig:SNA399}}
\end{center}
\end{figure}

Abell399\_3\_14\_0 was discovered November 6, 2008 (preimaging with the
2.3 m Bok telscope on October 25, 2008 also caught the SN on the rise).
As can be seen in the left panel of Figure~\ref{fig:SNA399}, the SN
was relatively near a cluster red sequence member, which we will refer
to as Galaxy 1.  Nonetheless, our deep stack image at the time indicated
that the SN was $\sim$7 $R_{e}$ from the potential host's center, and so
we triggered our Gemini ToO observations.  Further analysis of the field
indicates that the SN did occur $>7 R_{e}$ from the galaxy center based
on SExtractor's {\sc phot\_fluxfrac} parameter.

To investigate further, we used GALFIT \citep{galfit} to fit
parameterized models to Galaxy 1.  Several simple models with either a
single de Vaucouleurs or Sersic profile, along with a central point
source, were fit to Galaxy 1, all suggesting that the SN was
$\gtrsim5.5$ effective radii from the galaxy center.  However, these
models had significant residuals, suggesting that an additional
component was necessary to fit the data.  After some investigation, the
best model consists of two Sersic components, along with a point source
at the galaxy's center.  The larger Sersic component has an effective
radius of 2.2 arcseconds (3.3 kpc), suggesting that Abell399\_3\_14\_0
was only $\sim$3.8 effective radii from the galaxy center, at least with
respect to this largest scale component.  Given the Sersic index of this
component, $n=7.7$, $\sim$17\% of the light lies at larger radii than
the SN radius.  In Figure~\ref{fig:SNA399}, middle panel, we show the
two-component Sersic model of Galaxy 1 subtracted from the image.  Note
that there are low surface brightness features projected at large radii
around Galaxy 1, possibly a plume or tidal debris, suggesting that it
has recently interacted with another galaxy.  Signs of disturbance and
tidal features are common in both field ellipticals
\citep{vandokkum05,Tal09}, and their cluster counterparts
\citep{Janowiecki10}.

To corroborate our GALFIT results, we implemented a non-parametric
approach to measuring the light profile of Galaxy 1.  The data around
Galaxy 1 were binned into 1 pixel-wide annuli, with deviant pixels
flagged and removed through an iterative procedure.  A sky annulus was
taken, 20-pixels wide, and this sky value was subtracted from Galaxy 1's
light profile.  This allowed for a direct measurement of the fraction of
light enclosed as a function of radius, with $\sim$10\% of Galaxy 1's
light at radii larger than that of the SN.  This rough agreement
between the GALFIT and non-parametric analyses indicate that
Abell399\_3\_14\_0 lies in the outskirts of Galaxy 1 in projection.

Given the association of Abell399\_3\_14\_0 with Galaxy 1's outer
halo, and the appearance that Galaxy 1 has recently had a significant
interaction, it is interesting to look at the relative velocities of
Galaxy 1 (which was placed in the Gemini slit) and the SN. Precision
redshifts of SNe are difficult to measure due to their broad,
evolving spectral features.  That said, taking the median redshift of
the good correlations found by SNID -- based on 20 normal SN Ia
templates -- gives a redshift of $z=0.0603\pm0.0012$ (where the
uncertainty is the standard deviation of the good correlation
redshifts).

SNID also has several galaxy templates that it can use to quantify
galaxy redshifts, but since it smooths the data, we choose instead to
use the {\sc IRAF} task {\sc RVSAO} \citep{rvsao}, which is based on
the cross-correlation algorithm of \citet{Tonry79}, to precisely
measure the redshift of Galaxy 1 (note that SNID gives consistent
results with what follows, albeit with higher error bars).  Several
templates were used, based on absorption line spectra of nearby
galaxies along with a template made with K-giant spectra in M31
globular clusters.  All of these templates give nearly identical
results, to within $\delta z=0.0001$.  Taking the template which gave
the best r-statistic value \citep{Tonry79}, we measure a redshift of
$z=0.06439\pm0.00010$ for Galaxy 1.

While the velocity difference between Abell399\_3\_14\_0 and Galaxy 1
are formally more than 3-$\sigma$ apart, it is also insightful to plot
the velocity difference between Galaxy 1 and each of the good SN
template matches which SNID found, which we do in
Figure~\ref{fig:veldiffgal1}.  Note that all of the good SN template
matches indicate a velocity difference $>$500 km s$^{-1}$.  This
velocity difference suggests that Abell399\_3\_14\_0 is unbound to
Galaxy 1.  Assuming that Galaxy 1's half light radius is between
$\sim$1 kpc (our GALFIT result based on a single $R^{1/4}$ profile
fit) and $\sim$2 kpc (based on our nonparametric analysis of the light
profile), then we would expect Galaxy 1 to have a velocity dispersion
$\sim$100-200 km s$^{-1}$ \citep[e.g.][]{vanderWel08}.  Using this,
and the projected mass profile scaling relation of \citet{Bolton08}
based on strong gravitational lensing (which we deproject assuming
spherical symmetry), we find that Galaxy 1 would have an escape
velocity of $\sim$280-480 km/s at the radius of Abell399\_3\_14\_0,
presuming this is the true radius of the event.  We grant that this is
merely an estimate and that the application of the scaling relation of
\citet{Bolton08} requires extrapolation, but it is indicative that
Abell399\_3\_14\_0 is at best loosely bound to Galaxy 1.

We note that the redshift of Abell399\_3\_14\_0 derived via SNID puts
the SN just within our 3000 km s$^{-1}$ velocity cutoff for designating
cluster SNe.  To investigate if this velocity offset is reasonable, we
searched the NASA Extragalactic Database (NED) for redshifts in the
literature associated with Abell~399.  A standard biweight estimator
yields a cluster velocity dispersion of $\sigma=1230^{+80}_{-70}$ km
s$^{-1}$ from 123 cluster redshifts \citep[calculated as in][]{Just10},
putting Abell399\_3\_14\_0 in the 'outskirts' of the cluster in velocity
space -- but still within $\sim$2-3$\sigma$ given the SN's redshift
uncertainty.  With twenty-three cluster SN Ia in our sample, it is not
unreasonable to have a SN with such a velocity offset from the cluster.

The status of Abell399\_3\_14\_0 leans toward it being an IC SN.  In
projection, the SN lies in the outer regions of Galaxy 1, with
$\sim$10-20\% of the galaxy's light at larger radii than the SN.  On the
other hand, Galaxy 1 appears to have associated tidal debris and the
velocity offset between Galaxy 1 and the SN is larger than a simple
estimate of the escape velocity at the projected radius.  Indeed,
finding a true IC SNe projected onto a galaxy is {\it expected} for
samples roughly the size of MENeaCS (see \S~\ref{sec:bias}).  The bulk
of the evidence leads us to consider Abell399\_3\_14\_0 an IC object
based on these last facts.  However, if one wants to adopt the stricter
definition that IC stars are only those which are in the population
gravitationally dominated by the cluster rather than a local galaxy,
than Abell399\_3\_14\_0 could be a transitional object.  If we consider
the progenitor of Abell399\_3\_14\_0 to be part of a 'transition'
stellar population becoming unbound (or remaining loosely bound) from
its galaxy of birth, it does not appear to be alone in the
literature. \citet{Smith81} reported the discovery of SN 1980I, a SNIa,
to be projected onto a stellar bridge between NGC4374 and NGC4406 in the
Virgo cluster, suggesting a second 'transition' population object.  As
we calculate the IC stellar mass fraction in \S~\ref{sec:ICfrac}, we
will present our results both with and without Abell399\_3\_14\_0 as an
IC SN.

\section{Discussion \& Results}\label{sec:discuss}

In this section we will calculate the IC stellar mass fraction in
our sample of clusters based on the relative numbers of IC
and hosted SNe (\S~\ref{sec:ICfrac}) and discuss the relative
spatial distribution of these objects (\S~\ref{sec:spatial}).  First,
though, we quantify several observational biases which cause a
preference for the discovery and follow up of IC SNe
in MENeaCS (\S~\ref{sec:bias}).  Following this, we show that it is
unlikely that our IC SNe population are coming from dwarf
galaxies in the cluster, below our host detection limit, based on the
depth of our deep stack imaging fields and standard cluster galaxy
luminosity functions (\S~\ref{sec:hostlim}).  We discuss the
IC stellar population, with respect to the cluster galaxy
population as a whole in \S~\ref{sec:ICpop} with special emphasis on
what our SN sample contributes.

\subsection{Hostless supernova detection bias}\label{sec:bias}

There are two potential biases in using the relative numbers of hostless
and hosted cluster SNe to derive the IC stellar mass
fraction in our sample, which we attempt to quantify here.  These
assessments will be used in \S~\ref{sec:ICfrac} for properly calculating
our IC stellar mass fraction based on the relative numbers
of hostless and hosted cluster SN Ia, and we summarize them in
Table~\ref{table:ICbiastable}.  

First, it is simply more difficult to discover SNe within hosted
galaxies than it is in a seemingly blank portion of the sky.  This would
serve to boost the true number of hosted SNe relative to unhosted SNe.
To address this, we have carried out a series of detection efficiency
simulations by adding fake SN Ia into our MENeaCS images with realistic
absolute magnitude and stretch distributions using the template light
curves of \citet{Nugent02}.  One set of these fake SN Ia was implanted
with a distribution following the galaxy light in the image (out to a
maximum effective radius of 5$R_{e}$), while the other was purposely
placed at $R > 10 R_{e}$ away from any potential galaxy host in order to
mimic a hostless population (note that while this does not correspond to
our declared limit of $\gtrsim5 R_{e}$ for our IC SNe, the statistics
between implanted objects at $5 R_{e}$ and $> 10 R_{e}$ are identical).
The images, with their fake SN Ia implanted, were then run through our
detection pipeline and the statistics of the recovered SNe were tallied.
Details of this process will be presented when we calculate the cluster
SN Ia rate (Sand et al. in preparation), but we use the relative
detection efficiency between hosted and hostless SNe here to quantify
how much easier it was to discover hostless cluster SN Ia in the MENeaCS
survey.  The hosted to hostless SN detection efficiency is 0.91 when
integrated over all magnitudes and host-centric radii, and so a factor
of 1.0/0.91 will be applied to any value of the number of hosted cluster
SN~Ia, $N_{hosted}$, in relation to the number of IC SN~Ia, $N_{IC}$.

Second, there is a spectroscopic bias which always biases the number of
hosted SN~Ia to lower values.  Throughout the course of the MENeaCS
campaign, we made every effort possible to follow up {\it all}
transients that appeared to be hostless, as long as this was reasonable
to do within our Gemini ToO program or our classically scheduled
spectroscopic time.  Given our limited, classical spectroscopic
resources (which were effected by scheduling and weather) we only
followed up viable cluster SN Ia candidates with $g \lesssim 22.5$
during that time.  We split our spectroscopic bias into two categories,
and address each in turn.  First, there is what we call 'spectroscopic
availability bias' -- during the period in which our Gemini ToO program
was active we were always able to spectroscopically confirm hostless SN
candidates, even if we had no classical spectroscopic time due to
scheduling, weather or instrument problems.  Since we were able to get a
spectrum of all of our viable cluster SN Ia candidates as long as there
were no losses due to scheduling, weather, or technical problems, this
bias is easy to quantify.  It should simply be the ratio of time when
both classical spectroscopic time and Gemini ToO for hostless candidates
was available over the total amount of time that spectroscopic follow up
was available for hostless SN candidates -- we calculate this to be
0.91.  This means that another factor of 1.0/0.91 will be applied to any
value of $N_{hosted}$ when we calculate our IC stellar mass fraction.
As a sanity check, another way of quantifying the spectroscopic
availability bias is to directly investigate our transient database for
good cluster SN Ia candidates (as described in \S~\ref{sec:surveydesc})
that were not followed up due to poor weather or scheduling; three such
objects were found.  Based on the 0.91 fraction of classical
spectroscopic coverage found above, we would expect one or two such
objects -- perfectly consistent, especially given that good SN
candidates occasionally turned out to be background/foreground SNe or
even core collapse SNe.

The second type of spectroscopic bias is a 'hostless follow up' bias --
we were willing to spectroscopically confirm fainter, apparently
hostless, SN candidates with Gemini but would avoid pursuing similar
hosted candidates with our classical spectroscopic time.  The simplest
way of addressing this hostless follow up bias is to strike
Abell2495\_5\_13\_0 from our calculations of the IC stellar mass
fraction (\S~\ref{sec:ICfrac}).  With a discovery magnitude of $g=23.0$,
$r=22.4$, and declining light curve, such a cluster SN Ia caught so long
after maximum would not have been spectroscopically followed up and
confirmed if not for our Gemini ToO time.  All other IC SNe
would have been high priority targets in our classical spectroscopy
queue due to their magnitude and color.  

There is another bias, not considered in previous work, due to true IC
SNe being projected onto a false host galaxy, whether it be a
foreground/background galaxy or a cluster member.  To gauge the
potential magnitude of this 'filling factor' effect we calculated the
fractional image area enclosed within the dimensionless elliptical
radius $R < 5$ (see \S~\ref{sec:surveydesc}) of any potential host in a
representative sample of clusters, finding that on average $\sim$10 to
15\% of the image area is covered by a potential 'host galaxy'.
Additionally, when going from the cluster outskirts to the inner regions
of the cluster we find that the filled area increases by $\sim2-3$\%.
Of course, this filling factor estimation method assumes that IC SNe are
randomly distributed in the cluster, when in reality they could be more
strongly associated with galaxies if they have been recently stripped;
we do not quantify this plausible scenario further.  Our best estimate
suggests that $\lesssim$1 of our apparently hosted SN may be an actual
IC SN projected onto a galaxy's position.  One way to mitigate this
effect is to carefully compare SN and potential host redshifts (although
this is limited due to the broad, evolving features in SN spectra),
which we have done for the entire hosted SN set in the MENeaCS sample,
of which $\sim$30\% have a measured host redshift (Sand et al. in prep).
Besides Abell399\_3\_14\_0 and its large velocity offset with its
possible host galaxy (\S~\ref{sec:A399}), which we consider the first
object in this class, we have found no SN with significant velocity
offsets from their putative host.  Given the case of Abell399\_3\_14\_0,
we apply no additional correction factor since the expected size of the
effect is comparable in size to that observed. We gauge its impact by
presenting our results both with and without it in the IC sample.
Future work, in the era of large time domain surveys, will likely turn
up more such objects.

To summarize, we will apply three correction factors, based on
observational biases, to our IC stellar mass fraction analysis
in \S~\ref{sec:ICfrac}, which we show in Table~\ref{table:ICbiastable}.
First, we strike Abell2495\_5\_13\_0 from our IC stellar mass fraction
calculations for being too faint in comparison to our hosted SN~Ia.  We
then apply two separate factors of 1.0/0.91 whenever a $N_{hosted}$ term
is necessary: once for the relative difficulty of detecting hosted SNe
in comparison to hostless events, and once for the relative
spectroscopic availability of our IC SNe target of opportunity
observations and our classically scheduled spectroscopic time for which
we followed up our hosted SNe.

\subsection{Host limits and the cluster luminosity function}\label{sec:hostlim}

Given our host detection limits and a cluster galaxy luminosity
function, one can determine the fraction of the cluster's luminosity
contained in dwarfs fainter than our observational limits.  From this,
the probability of having the number of apparently hostless SNe in the
MENeaCS survey can be assessed, assuming that the SN~Ia rate is
roughly proportional to luminosity.

As a starting point, we choose the luminosity function and normalization
of the Virgo Cluster measured by \citet{Trentham02}, which has a
faint-end slope of $\alpha=-$1.03, and was measured down to $M_{R}=-10$
mag.  By integrating their luminosity function up to our detection limit
for faint hosts, and normalizing it by the total luminosity, the
fraction of cluster luminosity missed is calculated (not including the
contribution from IC stars), and we report this value for each IC field
in Table~\ref{table:deeptable}.  Note that in converting our reported
$M_{r}$ magnitudes in Table~\ref{table:deeptable} to the $M_{R}$
luminosity function used by \citet{Trentham02}, we made use of the
filter conversion relations of \citet{Blanton07} and the typical colors
of dwarf galaxies in SDSS reported by \citet{Eisenstein01}.

A variety of cluster luminosity function faint end slopes have been
measured in the literature -- for a recent compilation for the Coma
Cluster, see \citet{Milne07}.  The results of the last ten years
indicate that the Coma Cluster has a faint end slope between
$-1.5<\alpha<-1.0$.  We use a value of $\alpha=-$1.5 to give a realistic
upper limit on the fraction of the cluster luminosity in faint galaxies
below our deep stack detection limits, again using the normalization
terms of \citet{Trentham02}.  We list these values in
Table~\ref{table:deeptable}.

Assuming that the SN rate is proportional to luminosity, it is unlikely
that any of our IC SN candidates are actually hosted by faint dwarfs
below our detection limits.  If the typical faint end slope were
$\alpha=-1.03$, as measured for Virgo by \citet{Trentham02}, the typical
amount of galaxy light below our detection threshold is 0.04\%-0.12\% of
the total, suggesting that we would expect $\lesssim0.03$ SNe below our
detection limit.  Even given our upper limit scenario, with a faint end
slope of $\alpha=-1.5$ and our shallowest detection limit implies that
we would expect $\lesssim0.4$ SNe Ia hosted by galaxies below our
detection threshold.  A faint end slope of $\alpha\sim-1.85$ is
necessary for it to be plausible that all of our IC SN candidates were
hosted by dwarf galaxies below our detection limit.

We note that at some level all of these extrapolations are unphysical,
and should only be taken as indicative.  As is well known, a Schechter
function with a faint end slope steeper than $\alpha=-1.0$ has a
diverging number of dwarf galaxies, while faint end slopes steeper
than $\alpha=-2.0$ diverge in luminosity.  The faint end luminosity
function in clusters must truncate at the faintest luminosities,
strengthening our conclusion that our designated SNe are truly
IC events.

Of course, these strong constraints on faint dwarf galaxy hosts
implicitly assume that the SN~Ia rate is proportional to stellar
luminosity at the faintest luminosities.  Recent results from the Lick
Observatory Supernova Search (LOSS), however, suggest that there is a
strong dependence on the SN~Ia rate and the 'size' (and by extension,
luminosity) of a galaxy \citep{Li10}.  This rate-size relation indicates
that smaller galaxies have a higher SN rate per unit luminosity than
larger galaxies.  The physical mechanism for this effect is unknown,
although it may be due to a higher specific star formation rate among
smaller galaxies \citep{Li10}.  If true, then the rate-size relation is
not likely to hold in the cluster environment, since dwarf galaxies in
cluster cores have had their star formation shut off by environmental
processes \citep[e.g.][]{Penny10}.  We also note that the targeted
search strategy of LOSS did not include galaxies as faint as $M_{r}>-13$
mag, as we are discussing here.  Additionally, due to their extreme
stellar densities and low metallicities, globular clusters may have an
enhanced SN Ia rate per unit mass, perhaps as high as a factor of
$\sim$10 \citep{Shara02,Ivanova06,Rosswog08,Pfahl09} -- intracluster
globular clusters have been observed in the Virgo Cluster
\citep[e.g.][]{Lee10}.  Any effect of the rate-size relation on dwarf
galaxies below our detection limit would push our true IC stellar mass
fraction to lower values, as would any enhancement of the SN Ia rate in
globular clusters.  These possibilities make the case for even deeper
high-resolution imaging in the vicinity of our IC SN candidates very
exciting.

\subsection{What is the make-up of the intracluster stellar population?}\label{sec:ICpop}

A basic question when trying to use IC SNe (or for that matter, any
tracer of IC stars) as a measure of the IC stellar mass fraction is the
respective stellar populations of the cluster galaxies and the IC stars.
Much recent work has pointed out that the SN Ia rate of a stellar
population may have two components; one proportional to the stellar
mass, and the other to the star formation rate
\citep[e.g.][]{Mannucci05,Scannapieco05}.  Alternatively, there may be a
continuum of delay times between the birth of a stellar population and
when SN~Ia explosions occur \citep[e.g.][]{Pritchet08,Maoz10}.  If, for instance,
the IC stellar population was composed of solely old stars, while the
cluster galaxy population had a mix, including continuing small levels
of star formation, then the simple fraction of IC SNe out of the total
would not be a good measure of the IC stellar mass fraction.  Both the
observational and numerical picture of the IC stellar population are
varied and difficult to interpret, and we attempt to summarize the
current state of knowledge here.

Observationally, the most direct view of the ICL population has been in
Virgo via Hubble Space Telescope imaging of its red giant branch
population \citep{Williams07}.  Using color magnitude diagram matching
techniques, \citet{Williams07} concluded that the ICL population in
Virgo (at least in the field studied) is mostly old ($\gtrsim$10 Gyr)
with a wide range in metallicity ($-2.3 < [Fe/H] < 0.0$) but also has a
young, relatively more metal-rich component ($\lesssim$10 Gyr;
[Fe/H]$>0.5$) as well.  As a caveat, Virgo has a mass of
$\sim$1.7$\times 10^{14} M_{\odot}$ \citep{Rines06}, and is dynamically
unrelaxed \citep[e.g.][]{Binggeli85}, and so may not have ICL properties
analogous to the MENeaCS cluster sample, which has a median cluster mass
$\sim$3.8 times higher \citep[using the X-ray scaling relation
for the flux-limited sample of][and their bisector regression results]{Reiprich02}.  There is also observational evidence for current
star formation in IC space; the cold interstellar medium, removed from
its galaxy due to ram-pressure stripping, can form stars
\citep{Sun07,Sun10,Sivanandam10,Smith10}.  Broad band color imaging of the ICL, on
the whole, indicates that the ICL and the cluster galaxies have similar
stellar populations, with some studies finding colors that are more
similar to cluster galaxy outskirts, while others finding colors as red
as the interiors of early type galaxies; given the large uncertainties,
these color measurements are broadly consistent with the cluster galaxy
population \citep[e.g.][ among
others]{Gonzalez00,Zibetti05,Krick06,Krick07,daRocha08,Rudick10}.

Simulations have shown a similar range in results.  For instance,
simulations of ICL build up by \citet{Willman04} and \citet{Murante04}
have suggested that IC stars are more metal poor and older than the
general cluster galaxy population on average, respectfully.  Other
simulations indicate that IC stars have very similar properties to those
in cluster early type galaxies \citep{Sommer-Larsen05}.  Intriguingly,
and in line with some recent observations \citep{Sun10,Smith10},
\citet{Puchwein10} suggest that there may be some 'intracluster star
formation' coming from gas stripped from small infalling halos.  All of
the simulation results suffer from some overcooling, including those
with AGN feedback \citep[e.g.][]{Puchwein10}, and can not reproduce the
observed fraction of stars in cluster galaxies.  Thus simulation results
should be taken as only indicative.

A crude hint of the make-up of the ICL population can be gleaned from
the MENeaCS SN sample.  One of our IC SNe, Abell85\_6\_08\_0, has been
identified as a SN1991bg-like event, while none of our hosted SN Ia can
be placed in this class (although our sparse light curve sampling and
single spectrum per SN may preclude precise classification).
\citet{Li10} report that $\sim$18\% of all SNe Ia are of the
SN1991bg-like variety in a volume-limited survey, most appropriate for
our cluster SN search.  They prefer early type hosts \citep{Howell01},
and thus presumably are associated with an old stellar population
(although they may have a distribution of delay times).  Note that we
can not distinguish between scenarios in which the stellar population
was stripped while still young -- evolving passively in IC space since
-- or if the stellar population was stripped more recently with the old
stars developing in a galaxy evironment.  Given the small SN numbers
involved, the most robust conclusion is that there is an old stellar
population component to the IC light.  Future large time domain surveys
may be able to indirectly probe the stellar population of the ICL via
the mix of their SNe population.

As an aside, it should be mentioned that no IC core collapse SNe were
discovered in MENeaCS, while seven such hosted events were.  While the
focus of MENeaCS was on SNe Ia, we agressively followed up {\it all} of
our IC SN candidates.  In this situation, where hosted candidate core
collapse events were given lower spectroscopic priority but not their
hostless counterparts, we can place an upper limit of $f_{IC, CC} <
0.35$ (95\% confidence upper limit) on the fraction of IC core collapse
SNe.  This number was derived using binomial statistics
\citep{Gehrels86}, whose applicability is briefly discussed in
\S~\ref{sec:ICfrac}. If the core collapse SN rate is proportional to the
current star formation rate \citep[which it should be if core collapse
SN progenitors are from massive, short lived stars -- for a review
see][]{Galyam07}, then this fraction represents the relative star
formation rate in the IC population to that hosted by the cluster
galaxies.  While this constraint is not stringent, we again emphasize
that future work associated with the large time domain surveys will be
able to use IC SNe to probe the stellar population of the ICL.

\subsection{The intracluster stellar mass fraction}\label{sec:ICfrac}

The simplest assumption we can make is that the IC stellar population is
similar to that of the cluster galaxies.  Our presentation of the IC
stellar mass fraction in our sample will then focus on the fraction of
IC SNe to the total (corrected, of course, for the biases discussed in
\S~\ref{sec:bias}) as the proper tracer.  Nonetheless, we will also
present a conservative upper limit in case the IC stellar population is
more similar to that of the older, early type cluster galaxies
\citep[e.g.][]{Willman04,Murante04}, by utilizing those cluster SN Ia
hosted by red sequence galaxies (see below).  As we describe below,
binomial statistics are used to derive confidence limits, given that
intracluster and hosted SNe are each distinguishable, random events.  We
implicitly assume that the SN Ia rate is proportional to luminosity, and
any deviation from this among globular clusters and dwarf galaxies below
our faint host detection limit will push our intracluster stellar mass
fraction to lower values, as discussed in \S~\ref{sec:hostlim}.

In \S~\ref{sec:bias}, we quantified our known sources of bias associated
with discovery and confirmation of IC versus hosted cluster SN Ia
(summarized in Table~\ref{table:ICbiastable}).  Accordingly, two factors
of $1/0.91$ must be applied whenever the number of hosted cluster SN~Ia
is compared with the number of IC SNe, and Abell2495\_5\_13\_0 should be
struck from the sample since it would not have been followed up during
our normal, classically scheduled spectroscopic time.  This leaves three
IC SNe out of a total sample of twenty-two, keeping in mind that we will present
results with Abell399\_3\_14\_0 both in and out of the IC sample.  If we
wish to include only those cluster SNe within $R_{200}$ (which we
prefer, given that we begin becoming spectroscopically incomplete at
$R>1.2R_{200}$), where we have derived $R_{200}$ for each of the MENeaCS
clusters again using the X-ray scaling relations of \citet{Reiprich02},
then the total sample shrinks to sixteen.  We note that all but six of
the MENeaCS clusters have been searched for SNe out to $R_{200}$, given
the CFHT/Megacam field of view.  Even for these clusters, we were able
to cover out to $\sim0.9R_{200}$, and we thus apply no additional
correction factors to our SN numbers.

If the IC stellar population is exclusively old, we need a comparable SN
number among the hosted population in order to constrain the IC stellar
mass fraction.  To assess how many of our hosted cluster SN~Ia were
associated with old stellar regions, we have constructed red sequence
catalogs for each of our clusters (Bildfell et al. in prep), with a
typical scatter in the color magnitude relation of $\sim$0.03
magnitudes.  Seven (eight if one includes Abell399\_3\_14\_0 among the
hosted SNe) of the cluster SN Ia hosts are consistent with the red
sequence of their cluster (five are within $R_{200}$; six if you include
Abell399\_3\_14\_0).  We will take our red sequence galaxies to
represent an exclusively old stellar population, but acknowledge that
cluster red sequence galaxies may have an elevated SN Ia rate compared
to those in the field \citep{Sharon07,Mannucci08,Dilday10}, perhaps due
to residual star formation or a subtle metallicity effect.  We will
address this issue with MENeaCS data in future work, although all of the
appropriate studies suffer from small number statistics.  As
\citet{Rudnick09} measured $\sim$75\% of the clusters' total light
originating in red sequence members in their sample of SDSS clusters at
$\langle z \rangle=0.07$, and younger galaxies will have a higher mass
to light ratio, we conservatively assume that $\sim$75\% of our
clusters' stellar mass is in red sequence members in order to make a
correction to represent the clusters' old stellar population as a whole.
We apply this additional correction factor, which has the form $1/0.75$
whenever the number of SN~Ia hosted by cluster red sequence galaxies is
used to calculate our limits on the IC stellar mass fraction.  This
correction factor likely biases the IC stellar mass fraction high when
calculating a conservative upper limit for that quantity.

We list our IC stellar mass fraction, given our various scenarios, in
Table~\ref{table:ICLfractable}.  To be explicit, the IC stellar mass
fraction, $f_{IC,*}$, is found using

\begin{equation}\label{eqn:fic}
f_{IC,*} = \frac{N_{IC}}{N_{IC} + (N_{hosted} b_{obs} f_{pop})}
\end{equation}

\noindent where $N_{IC}$ is the number of IC SN Ia and $N_{hosted}$ is the
number of hosted SN Ia.  The quantity $b_{obs}$ is the total
observational bias factor determined in \S~\ref{sec:bias}, and is
$1/(0.91)^2$ throughout this work.  Finally, $f_{pop}$ is an additional
correction factor representing the fraction of the parent stellar
population that the hosted SN sample is probing.  The value of $f_{pop}$
is unity when we consider the IC stellar population and that of the
cluster galaxies to be identical, and $f_{pop}=1/0.75$ when we are
calculating the IC stellar mass fraction assuming that the IC stellar
population is exclusively old, and use the number of SN Ia hosted by red
sequence galaxies to represent the hosted old stellar population in the
clusters -- see the previous paragraph for details.  All of our
confidence limits were derived based on binomial statistics, since we
are measuring two different kinds of distinguishable, random events
(i.e.~intracluster and hosted SNe).  In particular, we use the
expressions of \citet{Gehrels86} for finding confidence limits on the IC
stellar mass fraction, $f_{IC,*}$.  We present all of our IC stellar
mass fraction numbers as 68\% confidence limits, while the red sequence
derived IC stellar mass fractions are 84\% one-sided confidence limits.

Focusing on our results within $R_{200}$, we find an ICL stellar mass
fraction of $0.16^{+0.13}_{-0.09}$ ($0.11^{+0.12}_{-0.07}$ if you do not
include Abell399\_3\_14\_0 as an IC SNe).  If we use our conservative
measure of the ICL stellar mass fraction (counting only our SNe Ia
hosted by red sequence galaxies and including Abell399\_3\_14\_0 among
our IC SNe), we derive an upper limit on the IC stellar mass fraction of
$<0.47$, which we consider the absolute upper range that our results
allow.

This IC stellar mass fraction is consistent with the bulk of the results
in the literature, which generally find $\sim$10-30\% at the cluster
scale \citep[e.g.][]{Galyam03,Gonzalez05,Zibetti05,Krick07}.
\citet{Gonzalez07} measured an ICL fraction versus cluster velocity
dispersion relation for a sample of 23 groups and clusters.  We apply
this scaling relation to our own sample to determine our expected ICL
fraction, correcting the Gonzalez et al. numbers for the available field
of view (using Figure 5 in Gonzalez et al. 2007), and the velocity
dispersion of the clusters in our sample (converting from the X-ray
luminosity using the relation of \citet{Wu99}).  Additionally, we apply
a rough correction factor of 0.8 to account for the fact that the
\citet{Gonzalez07} ICL scaling relation actually includes the BCG
luminosity, roughly $\sim$20\% of the total BCG+ICL luminosity budget
\citep[see][Figure 7]{Gonzalez05}.  We find an expected ICL fraction of
$\sim$0.19, which is consistent with our current, SN-based measurement.

This can be seen more clearly in Figure~\ref{fig:ICL_compare}, where we
compare our SN IC results with those at the group scale \citep{McGee10}
and the direct surface brightness measurements of
\citet{Gonzalez07}.  Two simple correction factors were made to the
original \citet{Gonzalez07} numbers for the purposes of this plot.
First, a factor of 0.8 was applied to account for the fact that the
Gonzalez et al. numbers were presented as a BCG + ICL fraction, rather
than the ICL alone; the factor of 0.8 is the rough fraction that
Gonzalez et al. found themselves as the typical fraction of the BCG +
ICL belonging to the ICL alone.  Second, we renormalize their ICL
numbers so that they reflect the fraction within $R<R_{200}$, rather
than the listed $R<R_{500}$ value (using Figure 5 of Gonzalez et
al. 2007).  We present two numbers for the McGee et al. intragroup SNe
results.  The lower, dashed box represents the raw McGee et
al. measurement of 19 intragroup SNe out of a total sample of 59.
The upper, dotted box is the final intragroup stellar mass fraction they
reported after assuming that the IC stars are universally old, the
hosted group stars contained a mix of stellar ages, and then applying a
correction factor to their raw SN numbers using the
two-component SN Ia rate measurement of \citet{Dilday10b}.  Note that
this should be analogous to our own estimation of the upper limit of the
IC stellar mass fraction, assuming that the IC stars are universally
old.  The velocity dispersion range of the McGee et al. group sample was
derived by taking their quoted mean group mass, with its standard
deviation, and converting to $\sigma$ using Equation 6 of
\citet{Yang07}.  Note that the \citet{Yang07} group catalog was the
origin of the McGee et al. group sample.

The MENeaCS box in Figure~\ref{fig:ICL_compare} represents our ICL
stellar mass fraction result (within $R_{200}$) including
Abell399\_3\_14\_0 as an IC SN.  The upper limit arrow represents the
limit on the IC stellar mass fraction assuming that all IC stars are
from an old stellar population (while including Abell399\_3\_14\_0 as an
IC SNe).  The combined SN results at the group scale with our own
confirm the declining IC stellar mass fraction, as a function of
cluster mass, seen by \citet{Gonzalez07}.

The decline in IC stellar mass fraction as a function of halo mass may
point to the assembly history of massive clusters.  Speaking in broad
terms, if one assumes that groups are the dominant building block of
clusters then the groups that made present day clusters could not have
had the high IC stellar mass fraction that we see today, unless some
mechanism could remove intragroup stars during the cluster formation
process.  Moderate redshift measurements of the intragroup stellar
fraction should be lower than those seen at low redshift to fit in with
expectations from hierarchical structure formation.

\subsection{The spatial distribution of intracluster supernovae}\label{sec:spatial}

In Figure~\ref{fig:SNspatial} we present the distribution of SNe Ia as a
function of projected distance from the cluster center, scaled by
$R_{200}$.  Represented by the black histogram, it is apparent that the
IC SNe are more centrally concentrated in the cluster than the general,
hosted population (keep in mind the MENeaCS SN coverage begins to be
incomplete at $R>R_{200}$) -- although this is subject to small number
statistics.  The fact that the ICL is more centrally concentrated than
the cluster's galaxies has been observed previously
\citep[e.g.][]{Zibetti05}.

The central concentration of the IC SNe is not a clear cut marker for
the underlying stellar population of the ICL.  On the one hand, this may
favor a scenario in which the IC stellar population is primarily old,
since low redshift cluster cores are dominated by red sequence galaxies
with old stellar populations \citep[e.g.][]{Urquhart10}.  On the other
hand, \citet{Smith10} have shown, in their study of ultraviolet tails in
the Coma cluster, that it is those few blue galaxies in Coma's central
regions (modulo projection effects) which are predominantly being
stripped and contributing young stars to IC space.  Our IC SNe
distribution can not distinguish between these two scenarios.

In a precursor cluster SN survey to the current work, \citet{SandSN08}
found 3 IC SN candidates at $R>R_{200}$ within their transient sample,
and none at smaller radii, suggesting a relative deficit if the
candidates proved authentic.  Although it is plausible that infalling
groups have their own detectable IC stellar population, this seems to be
at odds with the current work, which has a similar radial coverage.
Table~\ref{table:spectable} shows that our IC SNe were between $\sim$150
and 600 kpc from the cluster center (as measured from the BCG
position). We suggest that the IC SN candidates at high clustercentric
radii in \citet{SandSN08} were likely background events in need of
spectroscopic confirmation, although we can not rule out that individual
events were indeed associated with an extended IC stellar population.
Indeed, although not discussed earlier, MENeaCS had one viable IC SN Ia
candidate which was followed up with our Gemini ToO program which turned
out to be a background SN Ia at $z=0.2$.  Deep imaging may yet uncover a
faint host at its position, and other apparently hostless SNe are
routinely reported in the literature \citep{Sullivan06}.


\section{Summary and Conclusions}

The MENeaCS cluster SNe survey has spectroscopically confirmed a sample
of twenty-three cluster SN~Ia, four of which have no hosts upon close
inspection.  We use this sample of hostless SNe to constrain the IC
stellar mass fraction, and the extent of the ICL in clusters.  We have
presented confirmational spectroscopy of these IC SN candidates,
concluding that Abell85\_6\_08\_0 was likely a SN 1991bg-like event,
suggesting that a component of the IC stellar population is old, as seen
by others \citep[e.g.][]{Williams07}.  Deep CFHT/Megacam imaging
indicated that one of our events, Abell399\_3\_14\_0, is spatially
coincident with the outskirts of a nearby red sequence galaxy, but the
velocity offset between the SN and its putative host (along with evident
tidal debris in the region) show that it is likely unbound and thus an
IC SN.  We characterized our detection limit for faint host galaxies by
implanting artificial stars into our deep stack images and found
detection limits of $M_{g}\gtrsim-12.5$ and $M_{r}\gtrsim-13.0$ in all
cases.  Using the Virgo Cluster luminosity function of
\citet{Trentham02}, we calculate that $\lesssim0.1\%$ of the clusters'
luminosity is in dwarf galaxies below our detection limit.  Assuming a
very steep faint end slope of $\alpha=-1.5$ increases this number to
$\lesssim1.7\%$, making it very unlikely that more than one of our IC
SN Ia candidates was hosted by a faint dwarf, as long as there is no
mechanism for boosting the SN Ia rate among very faint hosts
\citep[e.g][]{Pfahl09,Li10}.  All four IC SNe occurred within $\sim$600
kpc of the cluster center, while our hosted SNe are seen out to
$R_{200}$ and beyond, illustrating that the ICL is more centrally
concentrated than the cluster galaxies.

After accounting for observational biases which made our IC SNe easier
to discover and follow up, we calculate an IC stellar mass fraction of
$0.16^{+0.13}_{-0.09}$ (68\% confidence limit) for all objects within
$R_{200}$, assuming that the cluster galaxies and the IC stars have a
similar mix of stellar populations.  If we instead assume that the IC
stellar population is exclusively old, and that the cluster galaxies
have a mix of stellar ages, we derive an upper limit on the IC stellar
mass fraction of $<0.47$ (84\% one-sided confidence limit).  This is in
excellent agreement with the majority of IC measurements in the
literature.  These results, along with the IC stellar mass fraction at
the group scale provided by the SNe study of \citet{McGee10}, allow us
to confirm the declining stellar mass fraction as a function of halo
mass reported by \citet{Gonzalez07}.  IC and intragroup
studies at higher redshift will further constrain the assembly history
of these massive halos.  Indeed, one exciting avenue is the
pursuit of IC SNe to higher redshift, where cosmological surface
brightness dimming will greatly reduce the utility of traditional direct
measurements of the ICL.


Future prospects for hostless SN studies are excellent.  With the advent
of large, dedicated time domain surveys such as the Palomar Transient
Factory, SkyMapper, Pan-Starrs, La Silla-QUEST Variability Survey and
the Large Synoptic Survey Telescope, the number of hostless SNe should
increase rapidly.  These additional numbers will provide the statistical
confidence necessary to truly constrain the extent and content of
diffuse light at a variety of halo mass scales.

\acknowledgments

Many thanks to Kathy Roth and Tom Matheson for ensuring that the Gemini
observations were taken properly.  We thank the CFHT queue observers and
software team for acquiring excellent data and making it available to us
in near real time.  We thank Yan Gao for making us aware of the
typographical error in Eqn. 1 of Sullivan et al. (2006). We are also
grateful to Stephenson Yang for his patience and diligence regarding
computer and network maintenance.  DJS thanks Nelson Caldwell for the
use of his spectral templates, along with Greg Rudnick and Tommaso Treu
for useful discussions.  DZ acknowledges support from NASA LTSA award
NNG05GE82G and NSF grant AST-0307492.  HH acknowledges support from a
Netherlands Organization for Scientific Research VIDI grant and a Marie
Curie International Reintegration Grant.  CP acknowledges the support of
the Natural Sciences and Engineering Research Council of Canada.  This
research has made use of the NASA/IPAC Extragalactic Database (NED)
which is operated by the Jet Propulsion Laboratory, California Institute
of Technology, under contract with the National Aeronautics and Space
Administration.  \bibliographystyle{fapj} \bibliography{apj-jour,mybib}

\begin{thebibliography}{83}
\expandafter\ifx\csname natexlab\endcsname\relax\def\natexlab#1{#1}\fi

\bibitem[{REV(????)}]{REVTEX41Control}
 ????

\bibitem[{08(1)}]{apsrev41Control}
08. 1

\bibitem[{{Abadi} {et~al.}(2006){Abadi}, {Navarro}, \& {Steinmetz}}]{Abadi06}
{Abadi}, M.~G., {Navarro}, J.~F., \& {Steinmetz}, M. 2006, \mnras, 365, 747

\bibitem[{{Arnaboldi} {et~al.}(2004){Arnaboldi}, {Gerhard}, {Aguerri},
  {Freeman}, {Napolitano}, {Okamura}, \& {Yasuda}}]{Arnaboldi04}
{Arnaboldi}, M., {Gerhard}, O., {Aguerri}, J.~A.~L., {Freeman}, K.~C.,
  {Napolitano}, N.~R., {Okamura}, S., \& {Yasuda}, N. 2004, \apjl, 614, L33

\bibitem[{{Barbary} {et~al.}(2010){Barbary}, {Aldering}, {Amanullah},
  {Brodwin}, {Connolly}, {Dawson}, {Doi}, {Eisenhardt}, {Faccioli}, {Fadeyev},
  {Fakhouri}, {Fruchter}, {Gilbank}, {Gladders}, {Goldhaber}, {Goobar},
  {Hattori}, {Hsiao}, {Huang}, {Ihara}, {Kashikawa}, {Koester}, {Konishi},
  {Kowalski}, {Lidman}, {Lubin}, {Meyers}, {Morokuma}, {Oda}, {Panagia},
  {Perlmutter}, {Postman}, {Ripoche}, {Rosati}, {Rubin}, {Schlegel},
  {Spadafora}, {Stanford}, {Strovink}, {Suzuki}, {Takanashi}, {Tokita},
  {Yasuda}, \& {Supernova Cosmology Project}}]{Barbary10}
{Barbary}, K., {et~al.} 2010, ArXiv e-prints, 1010.5786

\bibitem[{{Bertin} \& {Arnouts}(1996)}]{sexbib}
{Bertin}, E., \& {Arnouts}, S. 1996, \aaps, 117, 393

\bibitem[{{Binggeli} {et~al.}(1985){Binggeli}, {Sandage}, \&
  {Tammann}}]{Binggeli85}
{Binggeli}, B., {Sandage}, A., \& {Tammann}, G.~A. 1985, \aj, 90, 1681

\bibitem[{{Blanton} \& {Roweis}(2007)}]{Blanton07}
{Blanton}, M.~R., \& {Roweis}, S. 2007, \aj, 133, 734

\bibitem[{{Blondin} \& {Tonry}(2007)}]{SNID}
{Blondin}, S., \& {Tonry}, J.~L. 2007, \apj, 666, 1024

\bibitem[{{Bolton} {et~al.}(2008){Bolton}, {Treu}, {Koopmans}, {Gavazzi},
  {Moustakas}, {Burles}, {Schlegel}, \& {Wayth}}]{Bolton08}
{Bolton}, A.~S., {Treu}, T., {Koopmans}, L.~V.~E., {Gavazzi}, R., {Moustakas},
  L.~A., {Burles}, S., {Schlegel}, D.~J., \& {Wayth}, R. 2008, \apj, 684, 248

\bibitem[{{Da Rocha} {et~al.}(2008){Da Rocha}, {Ziegler}, \& {Mendes de
  Oliveira}}]{daRocha08}
{Da Rocha}, C., {Ziegler}, B.~L., \& {Mendes de Oliveira}, C. 2008, \mnras,
  388, 1433

\bibitem[{{Dilday} {et~al.}(2010{\natexlab{a}}){Dilday}, {Bassett}, {Becker},
  {Bender}, {Castander}, {Cinabro}, {Frieman}, {Galbany}, {Garnavich},
  {Goobar}, {Hopp}, {Ihara}, {Jha}, {Kessler}, {Lampeitl}, {Marriner},
  {Miquel}, {Moll{\'a}}, {Nichol}, {Nordin}, {Riess}, {Sako}, {Schneider},
  {Smith}, {Sollerman}, {Wheeler}, {{\"O}stman}, {Bizyaev}, {Brewington},
  {Malanushenko}, {Malanushenko}, {Oravetz}, {Pan}, {Simmons}, \&
  {Snedden}}]{Dilday10}
{Dilday}, B., {et~al.} 2010{\natexlab{a}}, \apj, 715, 1021

\bibitem[{{Dilday} {et~al.}(2010{\natexlab{b}}){Dilday}, {Smith}, {Bassett},
  {Becker}, {Bender}, {Castander}, {Cinabro}, {Filippenko}, {Frieman},
  {Galbany}, {Garnavich}, {Goobar}, {Hopp}, {Ihara}, {Jha}, {Kessler},
  {Lampeitl}, {Marriner}, {Miquel}, {Moll{\'a}}, {Nichol}, {Nordin}, {Riess},
  {Sako}, {Schneider}, {Sollerman}, {Wheeler}, {{\"O}stman}, {Bizyaev},
  {Brewington}, {Malanushenko}, {Malanushenko}, {Oravetz}, {Pan}, {Simmons}, \&
  {Snedden}}]{Dilday10b}
------. 2010{\natexlab{b}}, \apj, 713, 1026

\bibitem[{{Dolag} {et~al.}(2010){Dolag}, {Murante}, \& {Borgani}}]{Dolag10}
{Dolag}, K., {Murante}, G., \& {Borgani}, S. 2010, \mnras, 405, 1544

\bibitem[{{Eisenstein} {et~al.}(2001){Eisenstein}, {Annis}, {Gunn}, {Szalay},
  {Connolly}, {Nichol}, {Bahcall}, {Bernardi}, {Burles}, {Castander},
  {Fukugita}, {Hogg}, {Ivezi{\'c}}, {Knapp}, {Lupton}, {Narayanan}, {Postman},
  {Reichart}, {Richmond}, {Schneider}, {Schlegel}, {Strauss}, {SubbaRao},
  {Tucker}, {Vanden Berk}, {Vogeley}, {Weinberg}, \& {Yanny}}]{Eisenstein01}
{Eisenstein}, D.~J., {et~al.} 2001, \aj, 122, 2267

\bibitem[{{Fabricant} {et~al.}(2005){Fabricant}, {Fata}, {Roll}, {Hertz},
  {Caldwell}, {Gauron}, {Geary}, {McLeod}, {Szentgyorgyi}, {Zajac}, {Kurtz},
  {Barberis}, {Bergner}, {Brown}, {Conroy}, {Eng}, {Geller}, {Goddard},
  {Honsa}, {Mueller}, {Mink}, {Ordway}, {Tokarz}, {Woods}, {Wyatt}, {Epps}, \&
  {Dell'Antonio}}]{hectospec}
{Fabricant}, D., {et~al.} 2005, \pasp, 117, 1411

\bibitem[{{Foley} {et~al.}(2009){Foley}, {Matheson}, {Blondin}, {Chornock},
  {Silverman}, {Challis}, {Clocchiatti}, {Filippenko}, {Kirshner},
  {Leibundgut}, {Sollerman}, {Spyromilio}, {Tonry}, {Davis}, {Garnavich},
  {Jha}, {Krisciunas}, {Li}, {Pignata}, {Rest}, {Riess}, {Schmidt}, {Smith},
  {Stubbs}, {Tucker}, \& {Wood-Vasey}}]{Foley09}
{Foley}, R.~J., {et~al.} 2009, \aj, 137, 3731

\bibitem[{{Gal-Yam} {et~al.}(2007){Gal-Yam}, {Leonard}, {Fox}, {Cenko},
  {Soderberg}, {Moon}, {Sand}, {Li}, {Filippenko}, {Aldering}, \&
  {Copin}}]{Galyam07}
{Gal-Yam}, A., {et~al.} 2007, \apj, 656, 372

\bibitem[{{Gal-Yam} {et~al.}(2003){Gal-Yam}, {Maoz}, {Guhathakurta}, \&
  {Filippenko}}]{Galyam03}
{Gal-Yam}, A., {Maoz}, D., {Guhathakurta}, P., \& {Filippenko}, A.~V. 2003,
  \aj, 125, 1087

\bibitem[{{Gal-Yam} {et~al.}(2008){Gal-Yam}, {Maoz}, {Guhathakurta}, \&
  {Filippenko}}]{Galyam08}
------. 2008, \apj, 680, 550

\bibitem[{{Gal-Yam} \& {Simon}(2008)}]{GalyamIC2}
{Gal-Yam}, A., \& {Simon}, J. 2008, The Astronomer's Telegram, 1617, 1

\bibitem[{{Gehrels}(1986)}]{Gehrels86}
{Gehrels}, N. 1986, \apj, 303, 336

\bibitem[{{Germany}(1998)}]{GermanyIC}
{Germany}, L. 1998, \iaucirc, 6898, 1

\bibitem[{{Gonzalez} {et~al.}(2005){Gonzalez}, {Zabludoff}, \&
  {Zaritsky}}]{Gonzalez05}
{Gonzalez}, A.~H., {Zabludoff}, A.~I., \& {Zaritsky}, D. 2005, \apj, 618, 195

\bibitem[{{Gonzalez} {et~al.}(2000){Gonzalez}, {Zabludoff}, {Zaritsky}, \&
  {Dalcanton}}]{Gonzalez00}
{Gonzalez}, A.~H., {Zabludoff}, A.~I., {Zaritsky}, D., \& {Dalcanton}, J.~J.
  2000, \apj, 536, 561

\bibitem[{{Gonzalez} {et~al.}(2007){Gonzalez}, {Zaritsky}, \&
  {Zabludoff}}]{Gonzalez07}
{Gonzalez}, A.~H., {Zaritsky}, D., \& {Zabludoff}, A.~I. 2007, ArXiv e-prints,
  705, 0705.1726

\bibitem[{{Hamuy} {et~al.}(1995){Hamuy}, {Phillips}, {Maza}, {Suntzeff},
  {Schommer}, \& {Aviles}}]{Hamuy95}
{Hamuy}, M., {Phillips}, M.~M., {Maza}, J., {Suntzeff}, N.~B., {Schommer},
  R.~A., \& {Aviles}, R. 1995, \aj, 109, 1

\bibitem[{{Hook} {et~al.}(2004){Hook}, {J{\o}rgensen}, {Allington-Smith},
  {Davies}, {Metcalfe}, {Murowinski}, \& {Crampton}}]{Hook04}
{Hook}, I.~M., {J{\o}rgensen}, I., {Allington-Smith}, J.~R., {Davies}, R.~L.,
  {Metcalfe}, N., {Murowinski}, R.~G., \& {Crampton}, D. 2004, \pasp, 116, 425

\bibitem[{{Howell}(2001)}]{Howell01}
{Howell}, D.~A. 2001, \apjl, 554, L193

\bibitem[{{Ivanova} {et~al.}(2006){Ivanova}, {Heinke}, {Rasio}, {Taam},
  {Belczynski}, \& {Fregeau}}]{Ivanova06}
{Ivanova}, N., {Heinke}, C.~O., {Rasio}, F.~A., {Taam}, R.~E., {Belczynski},
  K., \& {Fregeau}, J. 2006, \mnras, 372, 1043

\bibitem[{{Janowiecki} {et~al.}(2010){Janowiecki}, {Mihos}, {Harding},
  {Feldmeier}, {Rudick}, \& {Morrison}}]{Janowiecki10}
{Janowiecki}, S., {Mihos}, J.~C., {Harding}, P., {Feldmeier}, J.~J., {Rudick},
  C., \& {Morrison}, H. 2010, ArXiv e-prints, 1004.1473

\bibitem[{{Just} {et~al.}(2010){Just}, {Zaritsky}, {Sand}, {Desai}, \&
  {Rudnick}}]{Just10}
{Just}, D.~W., {Zaritsky}, D., {Sand}, D.~J., {Desai}, V., \& {Rudnick}, G.
  2010, \apj, 711, 192

\bibitem[{{Komiyama} {et~al.}(2002){Komiyama}, {Sekiguchi}, {Kashikawa},
  {Yagi}, {Doi}, {Iye}, {Okamura}, {Shimasaku}, {Yasuda}, {Mobasher}, {Carter},
  {Bridges}, \& {Poggianti}}]{Komiyama02}
{Komiyama}, Y., {et~al.} 2002, \apjs, 138, 265

\bibitem[{{Krick} \& {Bernstein}(2007)}]{Krick07}
{Krick}, J.~E., \& {Bernstein}, R.~A. 2007, \aj, 134, 466

\bibitem[{{Krick} {et~al.}(2006){Krick}, {Bernstein}, \& {Pimbblet}}]{Krick06}
{Krick}, J.~E., {Bernstein}, R.~A., \& {Pimbblet}, K.~A. 2006, \aj, 131, 168

\bibitem[{{Kurtz} \& {Mink}(1998)}]{rvsao}
{Kurtz}, M.~J., \& {Mink}, D.~J. 1998, \pasp, 110, 934

\bibitem[{{Lee} {et~al.}(2010){Lee}, {Park}, \& {Hwang}}]{Lee10}
{Lee}, M.~G., {Park}, H.~S., \& {Hwang}, H.~S. 2010, Science, 328, 334

\bibitem[{{Li} {et~al.}(2010){Li}, {Leaman}, {Chornock}, {Filippenko},
  {Poznanski}, {Ganeshalingam}, {Wang}, {Modjaz}, {Jha}, {Foley}, \&
  {Smith}}]{Li10}
{Li}, W., {et~al.} 2010, ArXiv e-prints, 1006.4612

\bibitem[{{Mannucci} {et~al.}(2005){Mannucci}, {Della Valle}, {Panagia},
  {Cappellaro}, {Cresci}, {Maiolino}, {Petrosian}, \& {Turatto}}]{Mannucci05}
{Mannucci}, F., {Della Valle}, M., {Panagia}, N., {Cappellaro}, E., {Cresci},
  G., {Maiolino}, R., {Petrosian}, A., \& {Turatto}, M. 2005, \aap, 433, 807

\bibitem[{{Mannucci} {et~al.}(2008){Mannucci}, {Maoz}, {Sharon}, {Botticella},
  {Della Valle}, {Gal-Yam}, \& {Panagia}}]{Mannucci08}
{Mannucci}, F., {Maoz}, D., {Sharon}, K., {Botticella}, M.~T., {Della Valle},
  M., {Gal-Yam}, A., \& {Panagia}, N. 2008, \mnras, 383, 1121

\bibitem[{{Maoz} {et~al.}(2010){Maoz}, {Mannucci}, {Li}, {Filippenko}, {Della
  Valle}, \& {Panagia}}]{Maoz10}
{Maoz}, D., {Mannucci}, F., {Li}, W., {Filippenko}, A.~V., {Della Valle}, M.,
  \& {Panagia}, N. 2010, ArXiv e-prints, 1002.3056

\bibitem[{{McGee} \& {Balogh}(2010)}]{McGee10}
{McGee}, S.~L., \& {Balogh}, M.~L. 2010, \mnras, 403, L79

\bibitem[{{Milne} {et~al.}(2007){Milne}, {Pritchet}, {Poole}, {Gwyn},
  {Kavelaars}, {Harris}, \& {Hanes}}]{Milne07}
{Milne}, M.~L., {Pritchet}, C.~J., {Poole}, G.~B., {Gwyn}, S.~D.~J.,
  {Kavelaars}, J.~J., {Harris}, W.~E., \& {Hanes}, D.~A. 2007, \aj, 133, 177

\bibitem[{{Murante} {et~al.}(2004){Murante}, {Arnaboldi}, {Gerhard}, {Borgani},
  {Cheng}, {Diaferio}, {Dolag}, {Moscardini}, {Tormen}, {Tornatore}, \&
  {Tozzi}}]{Murante04}
{Murante}, G., {et~al.} 2004, \apjl, 607, L83

\bibitem[{{Nugent} {et~al.}(2002){Nugent}, {Kim}, \& {Perlmutter}}]{Nugent02}
{Nugent}, P., {Kim}, A., \& {Perlmutter}, S. 2002, \pasp, 114, 803

\bibitem[{{Peng} {et~al.}(2002){Peng}, {Ho}, {Impey}, \& {Rix}}]{galfit}
{Peng}, C.~Y., {Ho}, L.~C., {Impey}, C.~D., \& {Rix}, H. 2002, \aj, 124, 266

\bibitem[{{Penny} {et~al.}(2010){Penny}, {Conselice}, {de Rijcke}, {Held},
  {Gallagher}, \& {O'Connell}}]{Penny10}
{Penny}, S.~J., {Conselice}, C.~J., {de Rijcke}, S., {Held}, E.~V.,
  {Gallagher}, J.~S., \& {O'Connell}, R.~W. 2010, \mnras, 1429

\bibitem[{{Pfahl} {et~al.}(2009){Pfahl}, {Scannapieco}, \&
  {Bildsten}}]{Pfahl09}
{Pfahl}, E., {Scannapieco}, E., \& {Bildsten}, L. 2009, \apjl, 695, L111

\bibitem[{{Pritchet} {et~al.}(2008){Pritchet}, {Howell}, \&
  {Sullivan}}]{Pritchet08}
{Pritchet}, C.~J., {Howell}, D.~A., \& {Sullivan}, M. 2008, \apjl, 683, L25

\bibitem[{{Puchwein} {et~al.}(2010){Puchwein}, {Springel}, {Sijacki}, \&
  {Dolag}}]{Puchwein10}
{Puchwein}, E., {Springel}, V., {Sijacki}, D., \& {Dolag}, K. 2010, ArXiv
  e-prints, 1001.3018

\bibitem[{{Purcell} {et~al.}(2008){Purcell}, {Bullock}, \&
  {Zentner}}]{Purcell08}
{Purcell}, C.~W., {Bullock}, J.~S., \& {Zentner}, A.~R. 2008, \mnras, 391, 550

\bibitem[{{Reiprich} \& {B{\"o}hringer}(2002)}]{Reiprich02}
{Reiprich}, T.~H., \& {B{\"o}hringer}, H. 2002, \apj, 567, 716

\bibitem[{{Reiss} \& {Sabine}(1998)}]{ReissIC}
{Reiss}, D., \& {Sabine}, S. 1998, \iaucirc, 7015, 1

\bibitem[{{Rines} \& {Diaferio}(2006)}]{Rines06}
{Rines}, K., \& {Diaferio}, A. 2006, \aj, 132, 1275

\bibitem[{{Rosswog} {et~al.}(2008){Rosswog}, {Ramirez-Ruiz}, \&
  {Hix}}]{Rosswog08}
{Rosswog}, S., {Ramirez-Ruiz}, E., \& {Hix}, W.~R. 2008, \apj, 679, 1385

\bibitem[{{Rudick} {et~al.}(2010){Rudick}, {Mihos}, {Harding}, {Feldmeier},
  {Janowiecki}, \& {Morrison}}]{Rudick10}
{Rudick}, C.~S., {Mihos}, J.~C., {Harding}, P., {Feldmeier}, J.~J.,
  {Janowiecki}, S., \& {Morrison}, H.~L. 2010, ArXiv e-prints, 1003.4500

\bibitem[{{Rudnick} {et~al.}(2009){Rudnick}, {von der Linden}, {Pell{\'o}},
  {Arag{\'o}n-Salamanca}, {Marchesini}, {Clowe}, {De Lucia}, {Halliday},
  {Jablonka}, {Milvang-Jensen}, {Poggianti}, {Saglia}, {Simard}, {White}, \&
  {Zaritsky}}]{Rudnick09}
{Rudnick}, G., {et~al.} 2009, \apj, 700, 1559

\bibitem[{{Sand} {et~al.}(2008){Sand}, {Zaritsky}, {Herbert-Fort},
  {Sivanandam}, \& {Clowe}}]{SandSN08}
{Sand}, D.~J., {Zaritsky}, D., {Herbert-Fort}, S., {Sivanandam}, S., \&
  {Clowe}, D. 2008, \aj, 135, 1917

\bibitem[{{Scannapieco} \& {Bildsten}(2005)}]{Scannapieco05}
{Scannapieco}, E., \& {Bildsten}, L. 2005, \apjl, 629, L85

\bibitem[{{Shara} \& {Hurley}(2002)}]{Shara02}
{Shara}, M.~M., \& {Hurley}, J.~R. 2002, \apj, 571, 830

\bibitem[{{Sharon} {et~al.}(2010){Sharon}, {Gal-Yam}, {Maoz}, {Filippenko},
  {Foley}, {Silverman}, {Ebeling}, {Ma}, {Ofek}, {Kneib}, {Donahue}, {Ellis},
  {Freedman}, {Kirshner}, {Mulchaey}, {Sarajedini}, \& {Voit}}]{Sharon10}
{Sharon}, K., {et~al.} 2010, \apj, 718, 876

\bibitem[{{Sharon} {et~al.}(2007){Sharon}, {Gal-Yam}, {Maoz}, {Filippenko}, \&
  {Guhathakurta}}]{Sharon07}
{Sharon}, K., {Gal-Yam}, A., {Maoz}, D., {Filippenko}, A.~V., \&
  {Guhathakurta}, P. 2007, \apj, 660, 1165

\bibitem[{{Sivanandam} {et~al.}(2010){Sivanandam}, {Rieke}, \&
  {Rieke}}]{Sivanandam10}
{Sivanandam}, S., {Rieke}, M.~J., \& {Rieke}, G.~H. 2010, \apj, 717, 147

\bibitem[{{Sivanandam} {et~al.}(2009){Sivanandam}, {Zabludoff}, {Zaritsky},
  {Gonzalez}, \& {Kelson}}]{Sivanandam09}
{Sivanandam}, S., {Zabludoff}, A.~I., {Zaritsky}, D., {Gonzalez}, A.~H., \&
  {Kelson}, D.~D. 2009, \apj, 691, 1787

\bibitem[{{Smith}(1981)}]{Smith81}
{Smith}, H.~A. 1981, \aj, 86, 998

\bibitem[{{Smith} {et~al.}(2010){Smith}, {Lucey}, {Hammer}, {Hornschemeier},
  {Carter}, {Hudson}, {Marzke}, {Mouhcine}, {Eftekharzadeh}, {James},
  {Khosroshahi}, {Kourkchi}, \& {Karick}}]{Smith10}
{Smith}, R.~J., {et~al.} 2010, \mnras, 1238

\bibitem[{{Sommer-Larsen} {et~al.}(2005){Sommer-Larsen}, {Romeo}, \&
  {Portinari}}]{Sommer-Larsen05}
{Sommer-Larsen}, J., {Romeo}, A.~D., \& {Portinari}, L. 2005, \mnras, 357, 478

\bibitem[{{Sullivan} {et~al.}(2006){Sullivan}, {Le Borgne}, {Pritchet},
  {Hodsman}, {Neill}, {Howell}, {Carlberg}, {Astier}, {Aubourg}, {Balam},
  {Basa}, {Conley}, {Fabbro}, {Fouchez}, {Guy}, {Hook}, {Pain},
  {Palanque-Delabrouille}, {Perrett}, {Regnault}, {Rich}, {Taillet}, {Baumont},
  {Bronder}, {Ellis}, {Filiol}, {Lusset}, {Perlmutter}, {Ripoche}, \&
  {Tao}}]{Sullivan06}
{Sullivan}, M., {et~al.} 2006, \apj, 648, 868

\bibitem[{{Sun} {et~al.}(2010){Sun}, {Donahue}, {Roediger}, {Nulsen}, {Voit},
  {Sarazin}, {Forman}, \& {Jones}}]{Sun10}
{Sun}, M., {Donahue}, M., {Roediger}, E., {Nulsen}, P.~E.~J., {Voit}, G.~M.,
  {Sarazin}, C., {Forman}, W., \& {Jones}, C. 2010, \apj, 708, 946

\bibitem[{{Sun} {et~al.}(2007){Sun}, {Donahue}, \& {Voit}}]{Sun07}
{Sun}, M., {Donahue}, M., \& {Voit}, G.~M. 2007, ArXiv e-prints, 706, 0706.1220

\bibitem[{{Tal} {et~al.}(2009){Tal}, {van Dokkum}, {Nelan}, \&
  {Bezanson}}]{Tal09}
{Tal}, T., {van Dokkum}, P.~G., {Nelan}, J., \& {Bezanson}, R. 2009, \aj, 138,
  1417

\bibitem[{{Taubenberger} {et~al.}(2008){Taubenberger}, {Hachinger}, {Pignata},
  {Mazzali}, {Contreras}, {Valenti}, {Pastorello}, {Elias-Rosa},
  {B{\"a}rnbantner}, {Barwig}, {Benetti}, {Dolci}, {Fliri}, {Folatelli},
  {Freedman}, {Gonzalez}, {Hamuy}, {Krzeminski}, {Morrell}, {Navasardyan},
  {Persson}, {Phillips}, {Ries}, {Roth}, {Suntzeff}, {Turatto}, \&
  {Hillebrandt}}]{Tauenberger08}
{Taubenberger}, S., {et~al.} 2008, \mnras, 385, 75

\bibitem[{{Tonry} \& {Davis}(1979)}]{Tonry79}
{Tonry}, J., \& {Davis}, M. 1979, \aj, 84, 1511

\bibitem[{{Trentham} \& {Tully}(2002)}]{Trentham02}
{Trentham}, N., \& {Tully}, R.~B. 2002, \mnras, 335, 712

\bibitem[{{Urquhart} {et~al.}(2010){Urquhart}, {Willis}, {Hoekstra}, \&
  {Pierre}}]{Urquhart10}
{Urquhart}, S.~A., {Willis}, J.~P., {Hoekstra}, H., \& {Pierre}, M. 2010,
  \mnras, 406, 368

\bibitem[{{van der Wel} {et~al.}(2008){van der Wel}, {Holden}, {Zirm}, {Franx},
  {Rettura}, {Illingworth}, \& {Ford}}]{vanderWel08}
{van der Wel}, A., {Holden}, B.~P., {Zirm}, A.~W., {Franx}, M., {Rettura}, A.,
  {Illingworth}, G.~D., \& {Ford}, H.~C. 2008, \apj, 688, 48

\bibitem[{{van Dokkum}(2005)}]{vandokkum05}
{van Dokkum}, P.~G. 2005, \aj, 130, 2647

\bibitem[{{Williams} {et~al.}(2007){Williams}, {Ciardullo}, {Durrell},
  {Vinciguerra}, {Feldmeier}, {Jacoby}, {Sigurdsson}, {von Hippel}, {Ferguson},
  {Tanvir}, {Arnaboldi}, {Gerhard}, {Aguerri}, \& {Freeman}}]{Williams07}
{Williams}, B.~F., {et~al.} 2007, \apj, 656, 756

\bibitem[{{Williams} {et~al.}(2004){Williams}, {Olszewski}, {Lesser}, \&
  {Burge}}]{90prime}
{Williams}, G.~G., {Olszewski}, E., {Lesser}, M.~P., \& {Burge}, J.~H. 2004, in
  Ground-based Instrumentation for Astronomy. Edited by Alan F. M. Moorwood and
  Iye Masanori. Proceedings of the SPIE, Volume 5492, pp. 787-798 (2004)., ed.
  A.~F.~M. {Moorwood} \& M.~{Iye}, 787--798

\bibitem[{{Willman} {et~al.}(2004){Willman}, {Governato}, {Wadsley}, \&
  {Quinn}}]{Willman04}
{Willman}, B., {Governato}, F., {Wadsley}, J., \& {Quinn}, T. 2004, \mnras,
  355, 159

\bibitem[{{Wu} {et~al.}(1999){Wu}, {Xue}, \& {Fang}}]{Wu99}
{Wu}, X., {Xue}, Y., \& {Fang}, L. 1999, \apj, 524, 22

\bibitem[{{Yang} {et~al.}(2007){Yang}, {Mo}, {van den Bosch}, {Pasquali}, {Li},
  \& {Barden}}]{Yang07}
{Yang}, X., {Mo}, H.~J., {van den Bosch}, F.~C., {Pasquali}, A., {Li}, C., \&
  {Barden}, M. 2007, \apj, 671, 153

\bibitem[{{Zibetti} {et~al.}(2005){Zibetti}, {White}, {Schneider}, \&
  {Brinkmann}}]{Zibetti05}
{Zibetti}, S., {White}, S.~D.~M., {Schneider}, D.~P., \& {Brinkmann}, J. 2005,
  \mnras, 358, 949

\end{thebibliography}

\clearpage

\begin{inlinefigure}
\begin{center}
\resizebox{\textwidth}{!}{\includegraphics{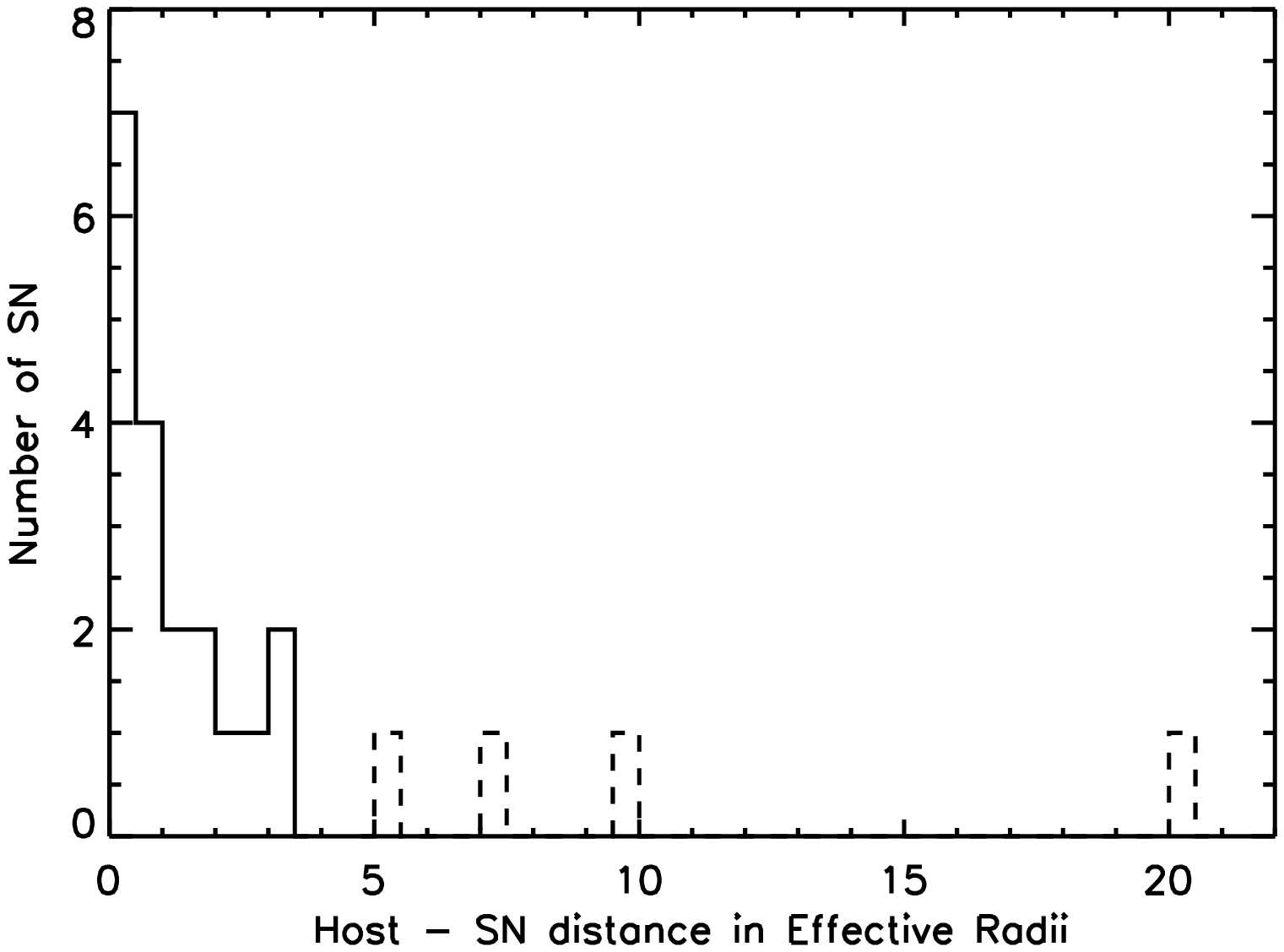}}
\end{center}
\figcaption{A histogram of the SN to host galaxy distance for the
cluster SNe Ia in the MENeaCS sample, expressed in units of the host's
effective radius, $R_{e}$.  The four IC SNe discussed in the current
work, shown as dashed lines, lie at a host -- SN distance $>5 R_{e}$
from the nearest possible host.  \label{fig:host_SN_dist}}
\end{inlinefigure}

\clearpage

\clearpage

\clearpage

\clearpage

\clearpage

\begin{inlinefigure}
\begin{center}
\resizebox{\textwidth}{!}{\includegraphics{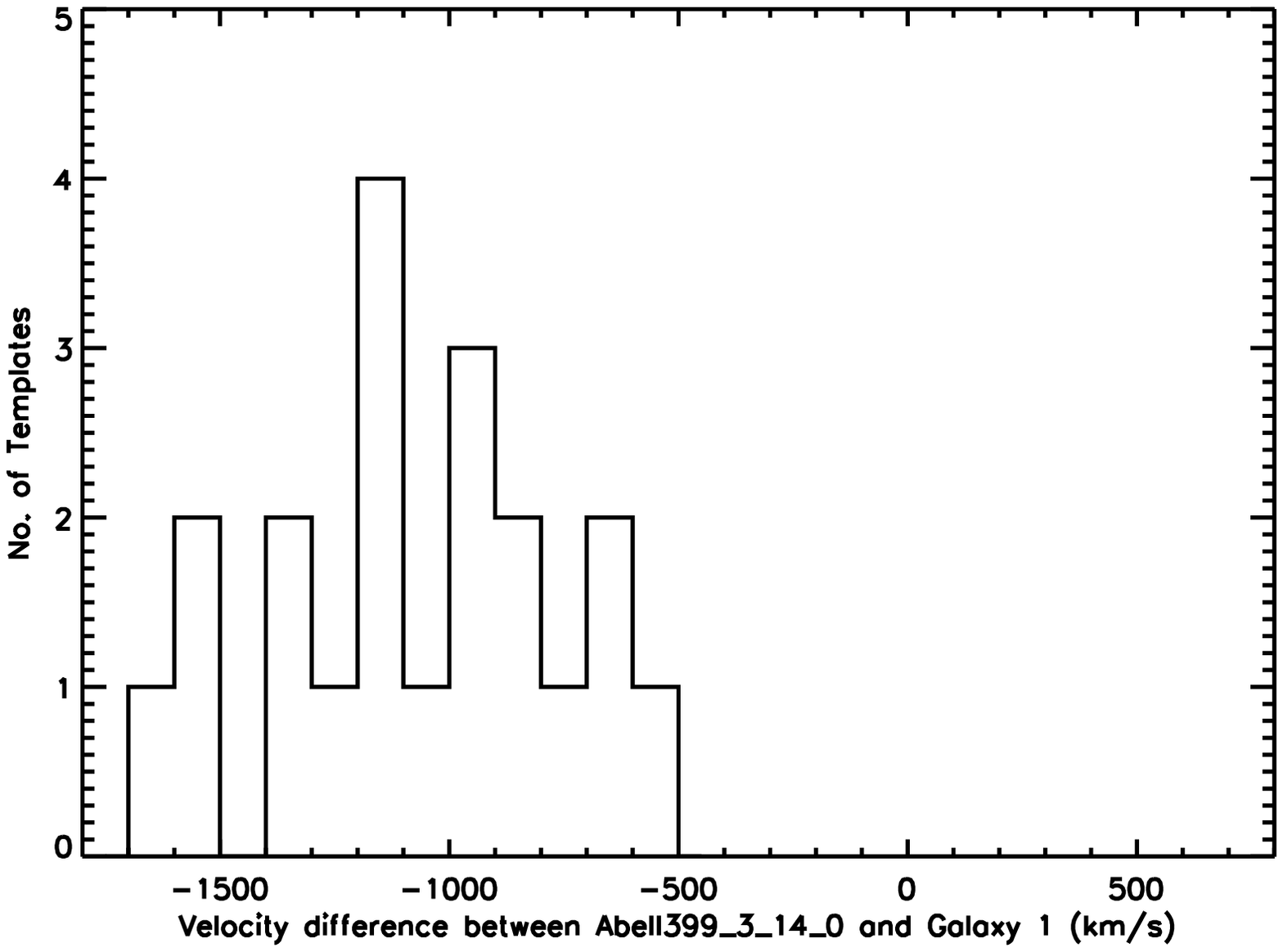}}
\end{center}
\figcaption{A histogram of the velocity offset between
Abell399\_3\_14\_0 and Galaxy 1 for the twenty good SNID template
matches to the SN.  None of the template redshifts are within 500 km
s$^{-1}$ of the galaxy.  We argue that Abell399\_3\_14\_0 is a part of a
stellar population which either is unbound or is becoming IC
stars since it lies in the outskirts of Galaxy 1, has a significant
velocity difference with that system, and Galaxy 1 itself has associated
low surface brightness plumes indicating a recent interaction (see
discussion in \S~\ref{sec:A399}).  \label{fig:veldiffgal1}}
\end{inlinefigure}

\clearpage

\begin{inlinefigure}
\begin{center}
\resizebox{\textwidth}{!}{\includegraphics{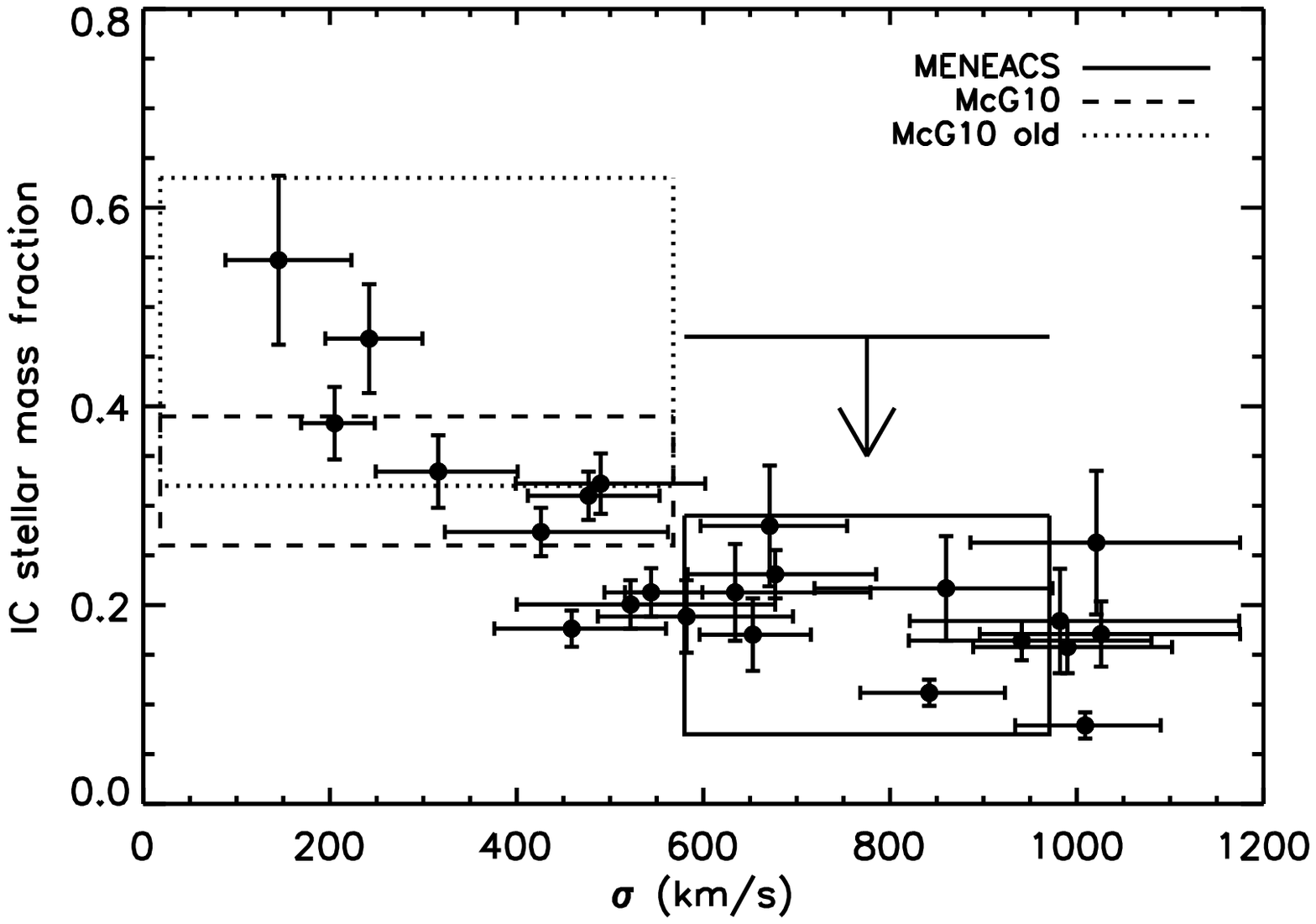}}
\end{center}
\figcaption{A comparison of the IC stellar mass fraction measured in the
current work (solid lined box for our numbers at $R<R_{200}$ including
Abell399\_3\_14\_0 as an IC SNe; the upper limit assumes that the IC
stars are universally old) with the results of \citet{Gonzalez07}, who
measured the ICL fraction via direct surface brightness observations for
a sample of 23 low redshift galaxy clusters (points with error bars).
Also plotted are the intragroup SN results of \citet{McGee10},
using both their raw numbers (dashed box) and those made after assuming
that the IC stellar population is composed of only old stars (dotted box).
The combined SN results confirm the declining IC stellar mass
fraction as a function of cluster mass seen by \citet{Gonzalez07}.
\label{fig:ICL_compare}}
\end{inlinefigure}

\clearpage

\begin{inlinefigure}
\begin{center}
\resizebox{\textwidth}{!}{\includegraphics{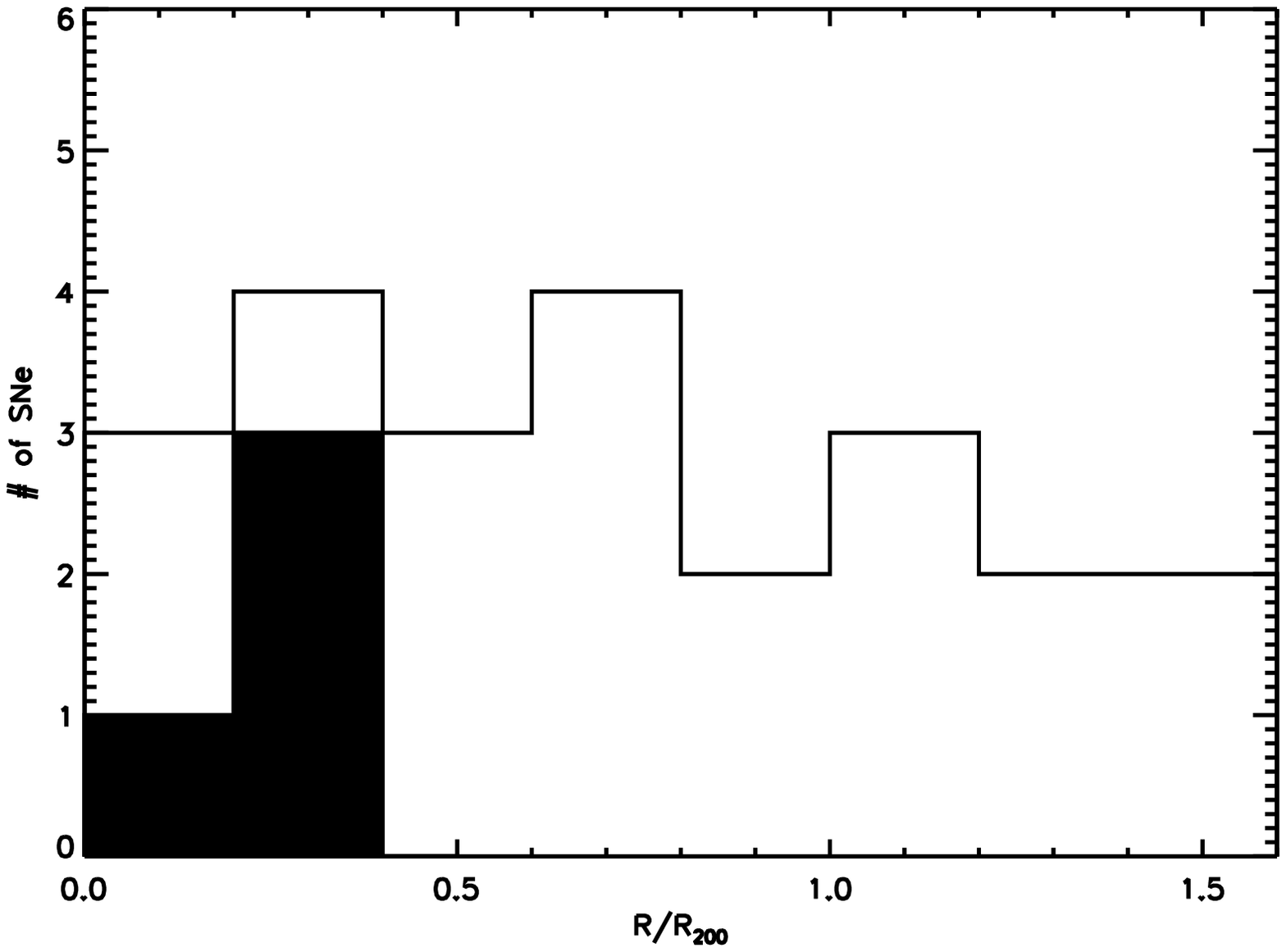}}
\end{center}
\figcaption{The distribution of cluster SN Ia in the MENeaCS survey,
with the IC SNe marked by the dark histogram.  The
IC SNe are more centrally concentrated than the
hosted cluster SN~Ia population.
\label{fig:SNspatial}}
\end{inlinefigure}

\clearpage

\begin{deluxetable}{lcccccccccc}
\tabletypesize{\scriptsize}
\tablecolumns{10}
\tablewidth{-100pt}
\tablecaption{MENeaCS intracluster supernovae \label{table:spectable}}
\tablehead{
\colhead{MENeaCS ID}  & \colhead{IAUC ID} & \colhead{UT Date} & \colhead{Telescope/} & \colhead{Type} & \colhead{$z_{SNID}$} & \colhead{Template} & \colhead{Phase} &\colhead{Exp. Time} & \colhead{Clustercentric}\\
\colhead{} & \colhead{} & \colhead{}
 &\colhead{Instrument}&\colhead{}&\colhead{}&\colhead{}&\colhead{(days)}&\colhead{(s)}
 &\colhead{Radius (kpc)}}\\
\startdata
Abell1650\_9\_13\_0 & ... & 2009-12-21.45 & MMT/Hecto & Ia-norm  & 0.0836 (0.0048) & SN2002er & -0.8 (4.7) & 2700.0 & 468 \\
Abell2495\_5\_13\_0 & SN2009hc & 2009-06-18.58& Gemini/GMOS & Ia-norm & 0.0796 (0.0032) & SN1998aq & 83.3 (15.1) & 3000.0 & 148\\
Abell85\_6\_08\_0  & ...  & 2009-07-04.60& Gemini/GMOS & Ia-91bg & 0.0617 (0.0007) & SN1991bg & 34.1 (5.5) & 1800.0 & 595 \\
Abell399\_3\_14\_0  & SN2008ih & 2008-11-28.49& Gemini/GMOS & Ia-norm  & 0.0613 (0.0025) & SN2003du & 14.4 (2.4) & 1200.0 & 616 \\
\enddata

\end{deluxetable}

\clearpage
\begin{deluxetable}{lcccccc}
\tablecolumns{7}
\tablecaption{Cluster Deep CFHT Stack Properties \label{table:deeptable}}
\tablehead{
\colhead{Cluster} &$z_{clus}$&\colhead{Exposure Time} & \colhead{$M_{g}$ limit} & \colhead{$M_{r}$ limit} & \colhead{$f(<L_{min})$} & \colhead{$f(<L_{min})$}\\
\colhead{} & \colhead{} & \colhead{(s)} & \colhead{(mag)} &\colhead{(mag)} &\colhead{$\alpha=-1.03$}&\colhead{$\alpha=-1.5$}}\\
\startdata
Abell~1650 & 0.0838 & 2280 & -12.47 & ... \\ 
& & 2240 & ... & -13.04 & 0.0012 & 0.0172\\
Abell~2495 & 0.0775 & 1920 & -11.72 & -12.37 & 0.0007 & 0.0127\\ 
Abell~85 & 0.0578 & 1920 & -11.15 & -11.68 & 0.0004 & 0.0091\\ 
Abell~399 & 0.0718 & 720 & -12.54 & ..\\  
& & 2400 & ... & -12.56 & 0.0009 & 0.0138 \\
 \enddata

\end{deluxetable}

\clearpage

\begin{deluxetable}{cc}
\tablecolumns{2}
\tablecaption{Intracluster SN bias correction factors from
 \S~\ref{sec:bias} \label{table:ICbiastable}}
\tablehead{
\colhead{Scenario} &\colhead{Correction Factor}  }\\
\startdata
Hosted to hostless SN detection efficiency  & 0.91\\
Spectroscopic availability bias  & 0.91 \\
Hostless follow-up bias  & strike Abell2495\_5\_13\_0 from IC calculations \\
 \enddata
\end{deluxetable}

\clearpage

\begin{deluxetable}{lcccccccccc}
\tablecolumns{10}
\tablewidth{0pc}
\tablecaption{Intracluster stellar mass fraction \label{table:ICLfractable}}
\tablehead{
\colhead{Scenario} &\colhead{Num. of IC SNe\tablenotemark{a}} & \colhead{Num. of hosted
 SNe\tablenotemark{a}}&\colhead{IC stellar mass fraction} }\\
\startdata

All hosts  & 2 & 20 & $0.08^{+0.09}_{-0.05}$\\
All hosts (with Abell399\_3\_14\_0 as IC) & 3 & 19 & $0.12^{+0.10}_{-0.07}$\\
Red sequence hosts & 2 & 8 & $<0.29$\\
Red sequence hosts (with Abell399\_3\_14\_0)& 3 & 7 &$<0.37$\\
Hosts within $R<R_{200}$ & 2 & 14 & $0.11^{+0.12}_{-0.07}$\\
Hosts within $R<R_{200}$ (with Abell399\_3\_14\_0)& 3 & 13 & $0.16^{+0.13}_{-0.09}$\\
Red sequence hosts within $R<R_{200}$ & 2 & 6 & $<0.36$\\
Red sequence hosts within $R<R_{200}$ (with Abell399\_3\_14\_0)& 3 & 5 & $<0.47$\\
 \enddata \tablenotetext{a}{These are the raw number of SNe,
after removing Abell2495\_5\_13\_0 (see \S~\ref{sec:bias}).  The
correction factors described in \S~\ref{sec:bias} are applied before the
IC stellar mass fraction is calculated. See Equation~\ref{eqn:fic}. }
\end{deluxetable}

\end{document}